\documentclass[%
 reprint,
superscriptaddress,
 amsmath,amssymb,
 aps,
prd,
]{revtex4-2}

\usepackage{changes}

\usepackage{xcolor}
\usepackage{graphicx}
\usepackage{mathtools}
\usepackage{orcidlink}
\usepackage{dcolumn}
\usepackage{bm}

\hypersetup{
    colorlinks=true, 
    citecolor=blue, 
    linkcolor=red
}

\def \be{\begin{equation}}
\def \bea{\begin{eqnarray}}
\def \eea{\end{eqnarray}}
\def \ee{\end{equation}}
\def\no{\nonumber}

\def \Msun {M_{\odot}}
\def \lmbdGW {\lambda_{\rm GW}}
\def \a {\alpha}
\def \b {\beta}
\def \ap {\alpha'}
\def \bp {\beta'}
\begin{document}

\title{A product template bank to search for compact binary coalescences microlensed by isolated point mass lenses}

\author{Sudhir Gholap\,\orcidlink{0009-0009-4987-7114}}
    \email[sudhir.gholap@iucaa.in]{}

\author{Sanjeev V. Dhurandhar}
    \email[sanjeev@iucaa.in]{}

\author{Shasvath J. Kapadia\,\orcidlink{0000-0001-5318-1253}}
    \email[shasvath.kapadia@iucaa.in]{}

\affiliation{Inter University Centre for Astronomy and Astrophysics, Post Bag 4, Ganeshkhind, Pune 411007}

\date{\today} 

\begin{abstract}
Gravitational waves (GW) from stellar mass compact binary coalescences (CBCs) that encounter a lens whose Schwarzschild radius is comparable to the GW wavelength, will be microlensed, leading to frequency-dependent modulations in the observed signal. Neglecting such wave-optics effects in standard templated searches for CBCs can result in signal-to-noise ratio (SNR) losses as large as $30\%$. In this work, we present a prescription for constructing a template bank targeted at microlensed GWs. We consider CBCs microlensed by isolated point mass lenses, whose frequency-dependent amplification factor is known to have a closed form analytical expression. However, to construct the template bank, we show that, it suffices to use the geometric optics approximation, effectively modeling the microlensed signal as overlapping signals with time-delays of $\mathcal{O}(1-500)$ ms. Moreover, given that matched filtering is mostly sensitive to the phase, as opposed to the amplitude, of the GW signal, we use the phase of the amplification factor to compute the metric elements. The usual CBC parameter space is appended by two extra parameters: (i)  the time-delay ($t_{\mathrm{d}}$) between images and, (ii) the relative-magnification ($\mu_{\mathrm{r}}$). By identifying the region of the lensing parameter space where the fitting factor (FF) falls below 97\% due to wave-optics effects, we construct a template bank consisting of $\sim 4000$ templates covering the selected lensing region, independent of the values of the unlensed CBCs' parameters. Indeed, we show that the bank targeting microlensed CBCs is, to a very good approximation, a Cartesian product of the unlensed CBC bank, and the bank spanning the lensing parameters. We demonstrate the effectiveness of our method by evaluating fitting factors using both the unlensed CBC bank, and the product template bank, finding a reduction of up to a factor of $\sim 10$ in the mismatch, as well as consistency with the prescribed minimal match criterion of $0.97$ used in the construction of the banks.
\end{abstract}

\maketitle

\section{Introduction}\label{sec:intro}

The LIGO
Scientific, Virgo, and KAGRA Collaboration (LVK) \cite{LIGOScientific:2014pky, VIRGO:2014yos, KAGRA:2020tym} collaboration has confirmed the detection of $\sim 400$ compact binary coalescences (CBCs) from three completed observing runs and two parts of the ongoing fourth observing run \cite{ligo2026gwtc}. These CBCs were identified using templated searches \cite{cannon2021gstlal, allene2025mbta, dal2021real, chu2022spiir}, where a bank of synthetic waveforms (templates) are cross correlated with detector data \cite{BSD-1, BSD-2, Owen:1995tm, BSO}. These matched filter-based searches crucially depend on the accuracy with which the templates represent the underlying signal in the data. Current template banks \cite{cokelaer2007gravitational, harry2009stochastic, roy2017hybrid} use quasi-circular, dominant-mode-only waveforms, assuming that such templates adequately model the last seconds/minutes that the CBC spends in the frequency band of the LVK detector network. Nevertheless, it has already been shown in the literature that omitting additional physical effects, such as eccentricity, precession, and higher harmonics, could result in appreciable loss of signal-to-noise ratio (SNR), and consequently, sensitivity, thereby missing potentially interesting candidates worthy of follow-up analyses \cite{DalCanton2015, Bustillo2016, Bustillo2021, Dietrich2019, Ramos-Buades2020, Phukon2025, Chia2024, Mehta:2025jiq}. In this paper, we focus on one such physical effect, viz., gravitational lensing. 

Gravitational lensing of GWs occurs when GWs from stellar mass CBCs encounter large agglomerations matter \cite{SchneiderEhlersFalco1992}. Galaxy and cluster scale lenses produce strong lensing in the geometric regime \cite{Ng2018,
li2018gravitational, oguri2018effect, smith2018if, smith2020massively,Smith:2019qsv, robertson2020does, ryczanowski2020building}, where the wavelength of the passing GW signal is much smaller than the Schwarzschild length scale of the lens mass ($\lambda_{\mathrm{GW}} \ll 2\mathrm{G}M_{\mathrm{lens}}/\mathrm{c}^2 $). This results in temporally resolved copies of the same signal with identical phase evolution, but differing amplitudes and a constant relative phase shift called the Morse phase \cite{Dai:2017huk}. Several strong-lensing search pipelines have been constructed \cite{gholap2025chi, goyal2021rapid, Goyal2024, Haris:2018vmn, kopty2026optimal, chakraborty2024glance, Janquart:2022wxc, Janquart:2023osz, lo2023, barsode2025fast, Ezquiaga2023, Campailla2026, li2023}, and used to search for signatures of strong lensing in pairs of GW signals. To date, no unambiguous strong lensing detection has been reported \cite{LIGOScientific:2021izm, LIGOScientific:2023bwz, LIGOScientific:2025cwb}. 

In this work, we restrict our attention to the case when $\lambda_{\mathrm{GW}} \sim 2\mathrm{G}M_{\mathrm{lens}}/\mathrm{c}^2 $, resulting in the occurrence of wave-optics effects \cite{takahashi2003wave}. GWs from stellar mass CBCs microlensed by isolated intermediate mass black holes (IMBHs) are an important example of this case, given their potential detectability with the LVK detector network. For IMBH lens masses spanning $100 - 1000 ~ \Msun$,  $\lmbdGW$ ranges from $300 - 3000$ km, which corresponds to a frequency range of $100 - 1000$ Hz and therefore falls in the sensitive band of the detectors. Such microlensed signals undergo frequency-dependent modulations in both amplitude and phase, relative to the unperturbed quasi-circular CBC waveform, thus enabling their identification (see, e.g., \cite{cheung2021stellar}). 

Microlensed GWs are expected to play an important role in advancing our understanding of astrophysics, cosmology and fundamental physics. For example, their identification could reveal crucial information about the populations of isolated IMBHs and their host environments, especially if these microlensed signatures are embedded in strongly lensed pairs of GWs \cite{shan2025interference, Seo:2021psp}. On the other hand, even their non-detection has important implications on our understanding of the nature of dark matter, placing upper limits on the fraction of dark matter as compact objects \cite{Basak:2021ten}. More speculatively, microlensed GWs could also reveal the presence of additional ``hairs'', such as beyond-Kerr spins \cite{Prabhu:2025elp}, and charge \cite{Deka:2024ecp}, profoundly altering our understanding of the nature of gravity. 

It is therefore imperative to incorporate microlensing features into CBC templates without rendering the resulting parameter space computationally infeasible to search. Several unmodeled search methods for microlensing signatures have been proposed, based on cross-correlation techniques \cite{seo2025residual, chakraborty2025mu, chakraborty2025first, chakraborty2026first}. In this work, we present, for the first time, a prescription for constructing a \textit{product template bank} to search for microlensed CBCs. Assuming a point-mass lens model, we show that, to a very good approximation, the product template bank is a Cartesian product between the standard CBC bank consisting of quasi-circular templates, and a \textit{lensing template bank} with templates characterized by the redshifted lens mass $M_{\mathrm{Lz}} = M_{\mathrm{L}}(1+z_{\mathrm{L}})$ and the dimensionless impact parameter $y$, written in units of the Einstein radius. The templates in the lensing template bank are constructed assuming the geometric optics approximation. In essence, then, the product template bank can be thought of as a collection of templates constructed from temporally overlapping (interfering) pairs of images separated by time delays of $\mathcal{O}(1-500)$ ms. We find that this approximation is more than adequate at recovering microlensed CBCs that also exhibit additional wave-optics effects. 

To assess the performance of the product template bank, we first identify the domain in the lens parameter space where the lack of microlensed CBC templates results in a significant loss of SNR, assuming an O4-like noise power spectral density (PSD), as implemented in \texttt{PyCBC} \citep{pycbc}. We do so by computing the match ($\mathcal{M}$), a noise-weighted inner product measuring the similarity between the unlensed templates ($h_{\mathrm{UL}}^{\mathrm{t}}$) and the lensed signals ($h_{\mathrm{L}}$), sampled across $M_{\mathrm{Lz}} \in [10, 10^5] M_{\odot}$ and $y\in [10^{-2}, 2]$. We find that in the parameter space considered, the loss of sensitive volume can be as egregious as $\sim 65\%$, consistent with previous works \cite{chan2025detectability, mishra2024unveiling}. We then leverage the geometric optics approximation to extract the phase correction to the unlensed GW signal and use its analytic form to lay out templates in the transformed time-delay ($t_{\mathrm{d}}$) -- relative-magnification ($\mu_{\mathrm{r}}$) lens-parameter space. We show that adopting the product template bank reduces the mismatch, defined as $1 - \mathcal{M}$, by up to a factor of $\sim 10$, and ensures that the minimal match criterion of $\mathcal{M} = 0.97$ used to construct the bank is met across the entire parameter space considered. 

This paper is organized as follows: In Sec.~\ref {sec:motivation}, we briefly introduce microlensing of GWs and present match plots to highlight the importance of a template bank for microlensed signals. We also quantitatively explain the occurrence of high-mismatch regions where a lensing bank is most needed. In Sec.~\ref {sec:method}, we derive the expression for the lensing phase matrix assuming a point-mass lens model, and use the matrix elements to obtain the template parameters for the lensing template bank. Sec.~\ref {sec:results} demonstrates the considerable improvement in sensitivity by adopting the product template bank. We conclude in Sec.~\ref{sec:summary_outlook}.

\section{The Need for a lensing template bank}\label{sec:motivation}

In this section, we motivate the need for a template bank targeting microlensed GWs. We provide a brief primer on microlensing of GWs in Sec.~\ref{subsec:intro_to_microlensing}, and present a match study in Sec.~\ref{subsec:loss_of_snr}, between the microlensed signals and the unlensed templates, highlighting important regions in the $M_{\mathrm{Lz}}-y$ parameter space, where match values drop significantly. We then provide, in Sec.~\ref{subsec:mismatch_explained}, a quantitative explanation for why loss of search sensitivity is particularly egregious in those regions. 

\subsection{Brief introduction to the microlensing of GWs}\label{subsec:intro_to_microlensing}

\begin{figure*}[h!t]
    \centering
    \includegraphics[width=\textwidth]    {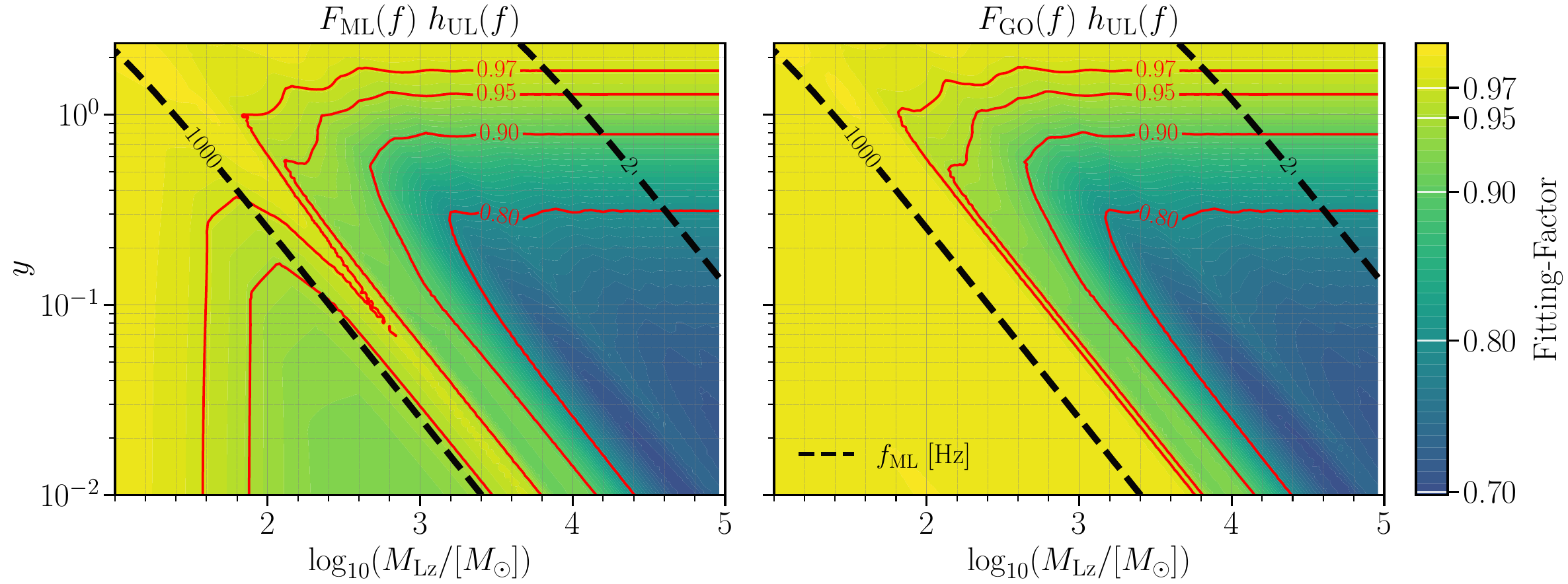}
    \caption{$FF$ computed for the lensed signals generated using Eq.~(\ref{eq:ff_microlens}) (left) and Eq.~(\ref{eq:geom_optics_ff}) (right), with the unlensed CBC templates generated using \textsc{IMRPhenomD} \cite{husa2016frequency, khan2016frequency} waveform, computed for a CBC signal of detector-frame total-mass $M_{\mathrm{tot}}=50 M_{\odot}$ and mass-ratio $q=1$. The solid-red lines show the fitting factor contours, while the black-dashed lines represent the constant $f_{\mathrm{ML}} \equiv 1 / t_{\mathrm{d}}$ contours. Both plots share a common feature of significantly low-$FF$ region, indicating that the geometric-optics approximation can be used as a reasonable approximation for approximating microlensed signals for practical purposes.}
\label{fig:mismatch_plot_IMR}
\end{figure*}

Consider a GW propagating in the spacetime curved by a lens mass $M_{\mathrm{L}}$. Assuming the weak-field limit ($|V/\mathrm{c}^2| \ll 1$, where $V$ is the gravitational potential) and thin-lens approximation, the unlensed signal ($h_{\mathrm{UL}}$) undergoes frequency-dependent modulations characterized by the lensing amplification factor ($F$), given by the Fresnel-Kirchhoff diffraction integral \cite{SchneiderEhlersFalco1992} as:
\begin{align}\label{eq:diffraction_integral}
    F(\omega, \mathbf{y}) = \frac{\omega}{2\pi i} \int d^2\mathbf{x} \, e^{i \omega \tau_{\mathrm{d}}(\mathbf{x}, \mathbf{y})},
\end{align}
where $\omega = \frac{8\pi \mathrm{G}M_{\mathrm{Lz}}f}{\mathrm{c}^3}$ is the dimensionless frequency and $M_{\mathrm{Lz}} = M_{\mathrm{L}}(1+z_{\mathrm{L}})$ is the redshifted lens mass. The vectors $\mathbf{x}$ and $\mathbf{y}$ denote the dimensionless image and source positions in the lens plane, respectively, expressed in units of the Einstein radius, $\xi^{}_{0} = \sqrt{\frac{D_L D_{LS}}{D_s} \frac{4GM_{Lz}}{c^2}}$. Here, $D_{\mathrm{S}}$, $D_{\mathrm{L}}$, and $D_{\mathrm{LS}}$ are the angular-diameter distances between the observer and source, observer and lens, and lens and source, respectively. The quantity $\tau_{\mathrm{d}}$ denotes the dimensionless time delay, given by
\begin{align}
    \tau_{\mathrm{d}}(\mathbf{x}, \mathbf{y}) = \frac{1}{2}|\mathbf{x}- \mathbf{y}|^2 - \psi_{\mathrm{L}}(\mathbf{x}) - \phi_{\mathrm{min}}(\mathbf{y}).
\end{align}
 The first term $\frac{1}{2} |\mathbf{x} - \mathbf{y}|^2$ accounts for the geometric time delay due to the extra path covered by the signal and the second term $\psi_{\mathrm{L}}(\mathbf{x})$ is the lensing potential that accounts for the Shapiro time delay. As is standard in the literature \cite{takahashi2003wave}, the term $\phi_{\mathrm{min}}(\mathbf{y})$ has been subtracted to ensure that the minimum of the time delay surface for a given $\mathbf{y}$ is zero.
\par

In the frequency domain, $h_{\mathrm{L}}$ is related to the $h_{\mathrm{UL}}$ as
\begin{equation}\label{eq:h_l}
    h_{\mathrm{L}}(f; \lambda_{\mathrm{L}}, \lambda_{\mathrm{CBC}}) = F(f; \lambda_{\mathrm{L}})\,h_{\mathrm{UL}}(f; \lambda_{\mathrm{CBC}}),
\end{equation}
where $\lambda_{\mathrm{L}}$ and $\lambda_{\mathrm{CBC}}$ are the lens and CBC parameters respectively. For the point-mass lens model, which adequately describes lensing by isolated compact objects such as IMBHs \cite{Lai2018}, the diffraction integral in Eq.~(\ref{eq:diffraction_integral}) is analytically solvable (see Appendix~\ref{subsec:microlens_ff_derivation}) and the exact expression is given by \cite{Deguchi1986, Nakamura1998, takahashi2003wave}
\begin{equation}\label{eq:ff_microlens}
\begin{split}
F_{\mathrm{ML}}(\omega, y) ={}&
\exp\Bigg[
    \frac{\omega \pi}{4}
    + \frac{i\omega}{2}
    \Bigg(
        \log\!\left(\frac{\omega}{2}\right)
        - 2\phi_{\mathrm{min}}(y)
    \Bigg)
\Bigg] \\
&\times
\Gamma\!\left( 1 - \frac{i\omega}{2} \right)\,
{}_1F_1\!\left(
    \frac{i\omega}{2},\, 1;\,
    \frac{i\omega y^2}{2}
\right).
\end{split}
\end{equation}
\begin{figure*}[h!t]
    \centering
    \includegraphics[width=\textwidth]{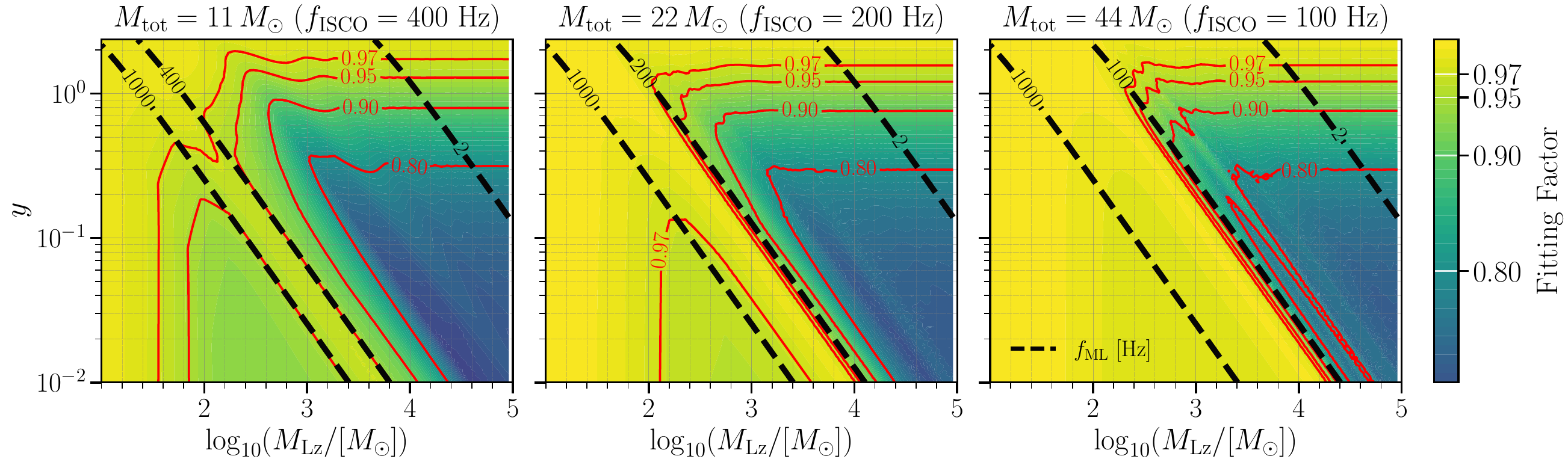}
    \caption{$FF$ computed for the microlensed signals generated using Eq.~(\ref{eq:ff_microlens}) with the non-spinning unlensed CBC templates generated using \textsc{TaylorF2} \cite{buonanno2009comparison} waveform, computed for a CBC signal of $M_{\mathrm{tot}}= \{11, 22, 44 \} M_{\odot}$ and mass-ratio $q=1$. The black dashed lines represent the constant $f_\mathrm{ML}$ contours. The low $FF$ region shrinks progressively as the $M_{\mathrm{tot}}$ is increased. The transition from high to low $FF$ happens around lens parameters, for which $f_{\mathrm{ML}} \approx f_{\mathrm{ISCO}}$.}
\label{fig:mismatch_plot_TF2}    
\end{figure*}
Here, the lensing amplification factor is characterised by parameters $\lambda_{\mathrm{L}} = \{ M_{\mathrm{Lz}}, y \}$, where $y = |\mathbf{y}|$, as the lens mass distribution has azimuthal symmetry. 
In the limit $\omega \to \infty$, the diffraction integral can be evaluated approximately using the stationary-phase approximation (see Appendix \ref{subsec:geom_ff_derivation}) to obtain the geometric-optics approximation for Eq.~(\ref{eq:diffraction_integral}), which is then given by \cite{Dai:2017huk},

\begin{equation}\label{eq:geom_optics_ff}
        F_{\mathrm{GO}}(f; M_{\mathrm{Lz}}, y) 
        = \sqrt{|\mu_{+}|} -i \sqrt{|\mu_{-}|}\mathrm{e}^{2\pi i f t_\mathrm{d}(M_{\mathrm{Lz}}, y)},
\end{equation}
where,
\begin{align}
    \mu_{\pm}(y) &= \frac{1}{2} \pm \frac{y^2 + 2}{2y\sqrt{y^2 + 4}},\\
    t_\mathrm{d}(M_{\mathrm{Lz}}, y) &= \frac{4 \mathrm{G} M_{\mathrm{Lz}}}{\mathrm{c}^3}\left[ \frac{y\sqrt{y^2 + 4}}{2} + \mathrm{ln}\left( \frac{y + \sqrt{y^2 + 4}}{-y + \sqrt{y^2 + 4}} \right) \right].
\end{align}
This effectively amounts to approximating the microlensed signal as a time-overlapping signal of two copies of $h_{\mathrm{{UL}}}$ with the amplitude of images scaled by $\sqrt{|\mu_{+}|}$ and $\sqrt{|\mu_{-}|}$, and having a time-delay between the images to be $t_{\mathrm{d}}$, which appears in the phase factor along with a constant Morse-phase shift of $-n\pi/2$ ($n = 0, 1)$ \cite{Dai:2017huk, Ezquiaga2021}.\\

\subsection{Loss of SNR due to inadequate modelling of microlensing effects}\label{subsec:loss_of_snr}

To understand the loss of SNR due to the absence of microlensing effects in the search pipelines, we present a match study between the microlensed signals ($h_{\mathrm{ML}}$), generated using $F(f) = F_{\mathrm{ML}}(f)$ in Eq.~(\ref{eq:ff_microlens}) and $h_{\mathrm{UL}}^{\mathrm{t}}$ from an unlensed template bank. For a detector characterized by a one-sided noise power spectral density (PSD) $S_{\mathrm{h}}(f)$, the noise-weighted complex inner product for waveforms $a$ and $b$ is defined as,
\begin{equation}
    (a|b) = 4 \int_{f_{\mathrm{min}}}^{f_\mathrm{\max}} \frac{a^{*}(f) b(f)}{S_{\mathrm{h}}(f)} df.
\end{equation}
For a given CBC signal, we sample the microlensed parameter space over a region $M_{\mathrm{Lz}} \in [10, 10^5] M_{\odot}$ and $y \in [0.01, 2]$ to generate $h_{\mathrm{ML}}$, and then compute the match $\mathcal{M}(h_{\mathrm{ML}}|h_{\mathrm{UL}}^{\mathrm{t}})$, maximized over an unlensed, non-spinning template parameter space, also commonly known as the fitting-factor. Here, $\mathcal{M}$ is defined as,
\begin{equation}\label{eq:match_ml_ul}
    \mathcal{M}(h_{\mathrm{ML}}|h_{\mathrm{UL}}^{\mathrm{t}}) = \mathrm{max}_{\{t_{\mathrm{c}}, \phi_{\mathrm{c}}\}} \left[ \frac{(h_{\mathrm{ML}}|h_{\mathrm{UL}}^{\mathrm{t}}(t_{\mathrm{c}}, \phi_{\mathrm{c}}))}{\sqrt{(h_{\mathrm{ML}}|h_{\mathrm{ML}})} \sqrt{(h_{\mathrm{UL}}^{\mathrm{t}}|h_{\mathrm{UL}}^{\mathrm{t}})}} \right],
\end{equation}
The maximization of the inner product $(h_{\mathrm{ML}}|h_{\mathrm{UL}}^{\mathrm{t}}(t_{\mathrm{c}}, \phi_{\mathrm{c}}))$ is performed over the coalescence time ($t_{\mathrm{c}}$) and coalescence phase ($\phi_{\mathrm{c}}$) of the unlensed template $h_{\mathrm{UL}}^{\mathrm{t}}$. The time maximization is efficiently computed via an inverse Fourier transform of the integrand, while the phase maximization is obtained by taking the absolute value of the complex inner product.
Without loss of generality, we have used the \textsc{IMRPhenomD} approximant to generate the CBC waveform \footnote{As will be clear in the forthcoming text, the prescription to construct the product template bank is independent of the nature of the templates and the physics that goes into their construction (e.g, spinning vs non-spinning, etc).}. The left panel of the Fig.~\ref{fig:mismatch_plot_IMR} shows the $FF$ computed for a lensed CBC having mass-ratio $q=1$ and detector-frame total-mass ($M_{\mathrm{tot}}$) of 50$M_{\odot}$.
We have also shown the $f_{\mathrm{ML}} \equiv 1 / t_{\mathrm{d}}$ contours (black dashed lines) in Fig. \ref{fig:mismatch_plot_IMR}.  In the small-$y$ limit, $t_{\mathrm{d}} \propto M_{\mathrm{Lz}}y$; consequently, contours of constant $f_{\mathrm{ML}}$ appear as straight lines with slope of $-1$ on a log–log scale. In the left panel of Fig.~\ref{fig:mismatch_plot_IMR}, we observe a loss in fitting-factor as severe as $30\%$ in the high-$M_{\mathrm{Lz}}$, low-$y$ region of the lens parameter space. For $ FF \sim 0.7$, the sensitive volume of the search pipeline reduces to $\sim FF^3 \approx 34\%$ of the total available volume. 
The right panel of the Fig.~\ref{fig:mismatch_plot_IMR} shows variation of $FF$ across the lens parameter space, when the lensed signals are generated using $F(f) =F_{\mathrm{GO}}(f)$ in the Eq.~(\ref{eq:h_l}). An important observation is that the region with severe loss of $FF \sim 0.7$ is common for the lensed signals under geometric-optics appoximation as well. This suggests that the geometric-optics approximation for the $F(f)$, given by Eq.~(\ref{eq:geom_optics_ff}) captures essential features of microlensing, which affect SNR the most, and thus, could be used for generating a lensed-template bank to improve the recovery of microlensed CBC signals.

\subsection{Low fitting-factor region explained}\label{subsec:mismatch_explained}

\begin{figure*}[t]
    \centering
    \includegraphics[width=\textwidth]{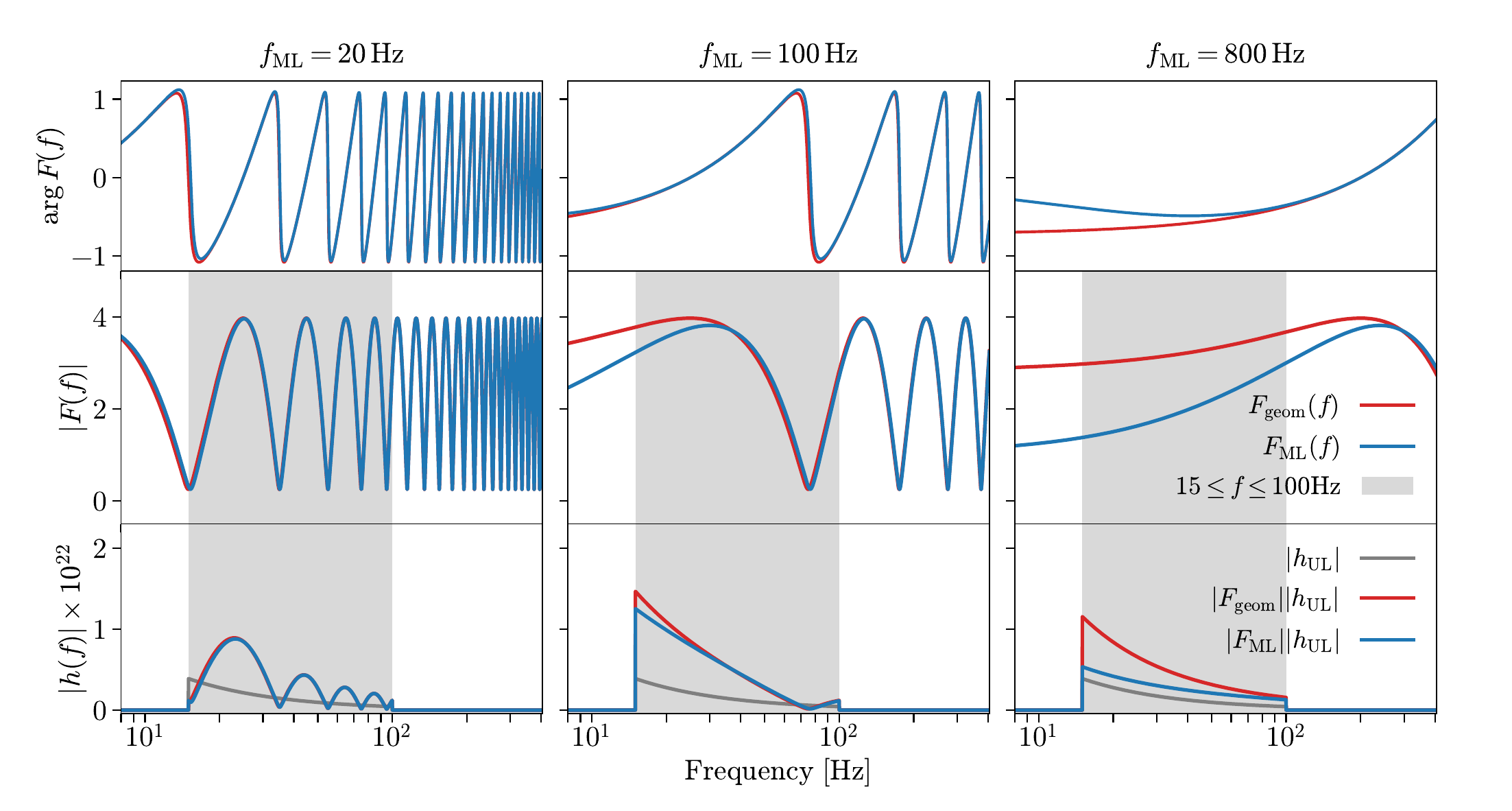}
    \caption{Phase (top panel) and amplitude (middle panel) of the amplification factors $F_{\mathrm{ML}}(f)$ and $F_{\mathrm{GO}}(f)$, along with their effect on the amplitude modulation of an inspiral-only signal (bottom panel) with total mass $M_{\mathrm{tot}} = 44\,M_{\odot}$ ($f_{\mathrm{ISCO}} = 100\,$Hz) and mass ratio $q = 1$. Results are shown for three lens configurations with $y = 0.127$ and $M_{\mathrm{Lz}} \in \{10^4,\, 2000,\, 250\}\,M_{\odot}$, corresponding to $f_{\mathrm{ML}} \in \{20,\, 100,\, 800\}$\,Hz, arranged from left to right. The gray shaded region indicates the signal frequency range.}
\label{fig:ff_comparison}    
\end{figure*}
\begin{figure}[]
    \makebox[\columnwidth]{%
        \hspace{2.95cm}
        \includegraphics[width=1.1\columnwidth, trim=15cm 2cm 2cm 1cm,clip]{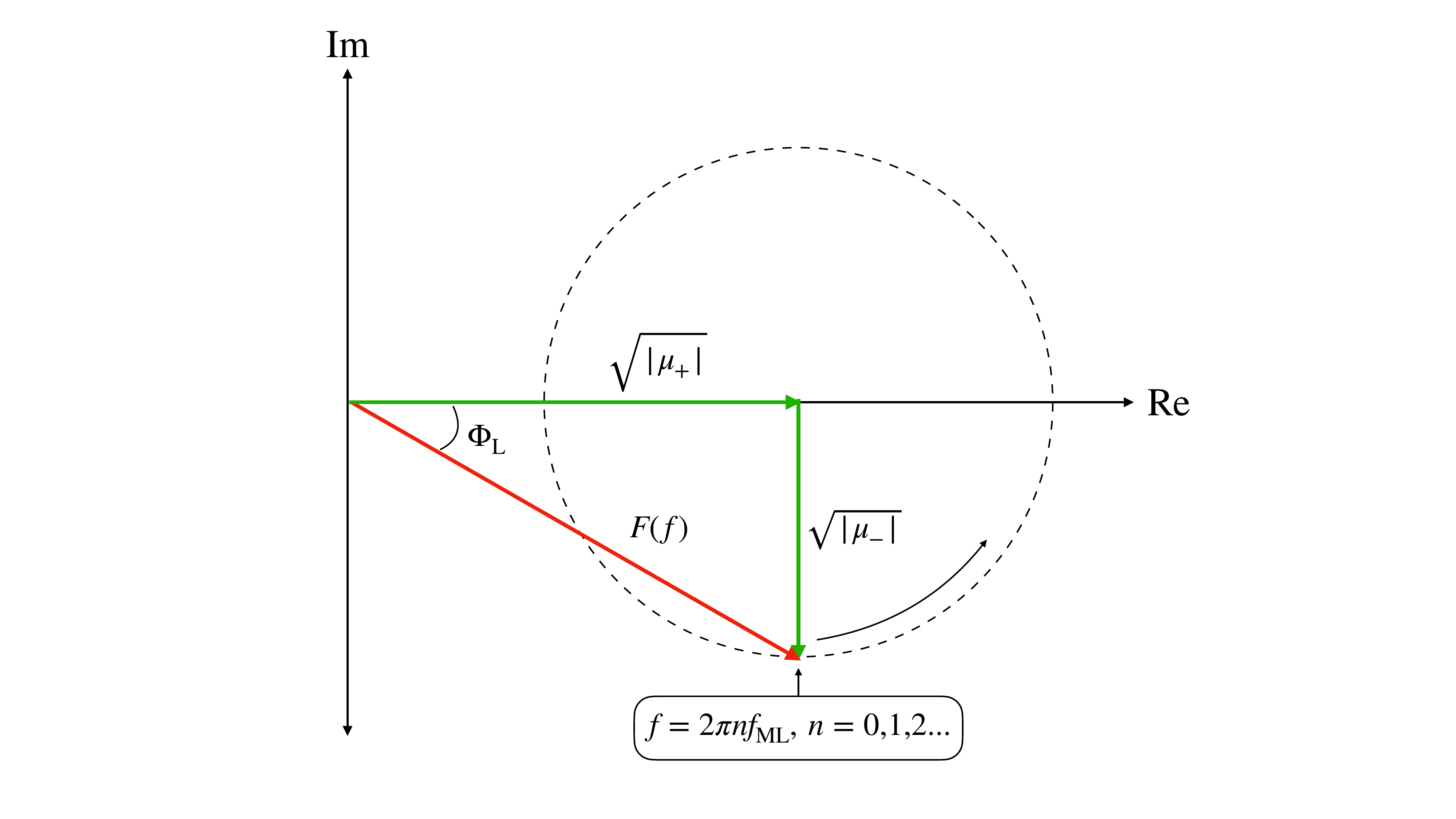}}
    \caption{A schematic for the frequency evolution of the $F_{\mathrm{GO}}(f)$ (shown in red), which is the sum of two complex numbers (shown in green), corresponding to the terms $\sqrt{|\mu_{+}|}$ and $-i\sqrt{|\mu_{-}|}\,\mathrm{e}^{2\pi if/f_{\mathrm{ML}}}$ in the complex plane.}
    \label{fig:ff_phasor}
\end{figure}

For better understanding, we compute $FF$ for different CBC signals having $q=1$ and $M_{\mathrm{tot}} \in \{ 11, 22, 44\} M_{\odot}$ respectively, with the CBC signal model being inspiral-only \textsc{TaylorF2}. As shown in Fig. \ref{fig:mismatch_plot_TF2}, a significant drop in the $FF$ is observed for parameters with high $M_{\mathrm{Lz}}$ and small $y$. This low-$FF$ region occurs for those lens-parameters, whose corresponding $f_{\mathrm{ML}}$ is less than the ISCO frequency of the CBC signal. It is also evident that the low-$FF$ region progressively shrinks as the CBC signal's $M_{\mathrm{tot}}$ increases. This is because higher-$M_{\mathrm{tot}}$ signals sweep through a shorter frequency range in-band. Consequently, such higher-mass systems are less susceptible to the periodic modulations imprinted by lensing in the frequency domain, compared to longer-duration, lower-mass systems. We illustrate this in Fig.~\ref{fig:ff_comparison}, which shows the amplitude and phase of $F(f)$ for three representative lens configurations. The top and middle panels display the phase ($\Phi_{\mathrm{L}}$) and amplitude of $F_{\mathrm{ML}}(f)$ and $F_{\mathrm{GO}}(f)$, respectively. The bottom panel shows the amplitude profile of an unlensed signal $h_{\mathrm{UL}}$ with $M_{\mathrm{tot}} = 44\,M_{\odot}$ and $q = 1$ ($f_{\mathrm{ISCO}}=100\,\mathrm{Hz}$), along with the cases where $|h_{\mathrm{UL}}|$ is modulated by $|F_{\mathrm{ML}}|$ and $|F_{\mathrm{GO}}|$. The grey shaded region indicates the signal frequency range $(15 \leq f \leq 100)\,\mathrm{Hz}$. The lens configurations are chosen to be $y = 0.127$ and $M \in \{10^4,\, 2000,\, 250\}\,M_{\odot}$, corresponding to $f_{\mathrm{ML}} \in \{20,\, 100,\, 800\}$ Hz, arranged from left to right. $f_{\mathrm{ML}}$ sets the scale for periodicity of amplitude and phase of $F_{\mathrm{GO}}(f)$ as illustrated by the phasor diagram in Fig.~\ref{fig:ff_phasor} for the geometric-optics approximation. The amplitude of $F_{\mathrm{GO}}(f)$ is oscillatory, and bounded between $(\sqrt{|\mu_{+}|} - \sqrt{|\mu_{-}|})$ and $(\sqrt{|\mu_{+}|} + \sqrt{|\mu_{+}|})$ with periodicity $f_{\mathrm{ML}}$. Similarly, the argument of $F_{\mathrm{GO}}(f)$ is also periodically varying as the frequency evolves, and is bounded between $\pm \pi/4$, which corresponds to the limiting case $\sqrt{|\mu_+|} \approx \sqrt{|\mu_-|}$, which happens when $y \rightarrow 0$. The left panel of Fig.~\ref{fig:ff_comparison} shows the case when $f_{\mathrm{ML}} < f_{\mathrm{ISCO}}$, which causes severe modulations in the amplitude and phase profile of the unlensed signal. In this configuration, the $F(f)$ is well approximated by the geometric optics approximation. The middle panel shows the case when $f_{\mathrm{ML}} = f_{\mathrm{ISCO}}$, which corresponds roughly to the transition region between high and low $FF$ regions in the right panel of Fig.~\ref{fig:mismatch_plot_TF2}. This demarcation of high to low $FF$ regions at $f_{\mathrm{ML}} = f_{\mathrm{ISCO}}$, is observed across all the CBC signals in Fig.~\ref{fig:mismatch_plot_TF2}. The rightmost panel of Fig.~\ref{fig:mismatch_plot_TF2} shows the case when $f_{\mathrm{ML}} \gg f_{\mathrm{ISCO}}$, where the microlensing causes an overall magnification of the unlensed signal and a little change to the phase. The geometric-optics approximation is no longer valid in this regime. Since there are no significant frequency-dependent modulations, this region is less affected by lensing and unlensed templates can recover signals getting microlensed from these parameters with $5\%$ loss of SNR at most, indicated by the lower-left region in the left panel of the Fig.~\ref{fig:mismatch_plot_IMR}. 
\par
\begin{figure}[]
    \centering
    \includegraphics[width=\columnwidth]    {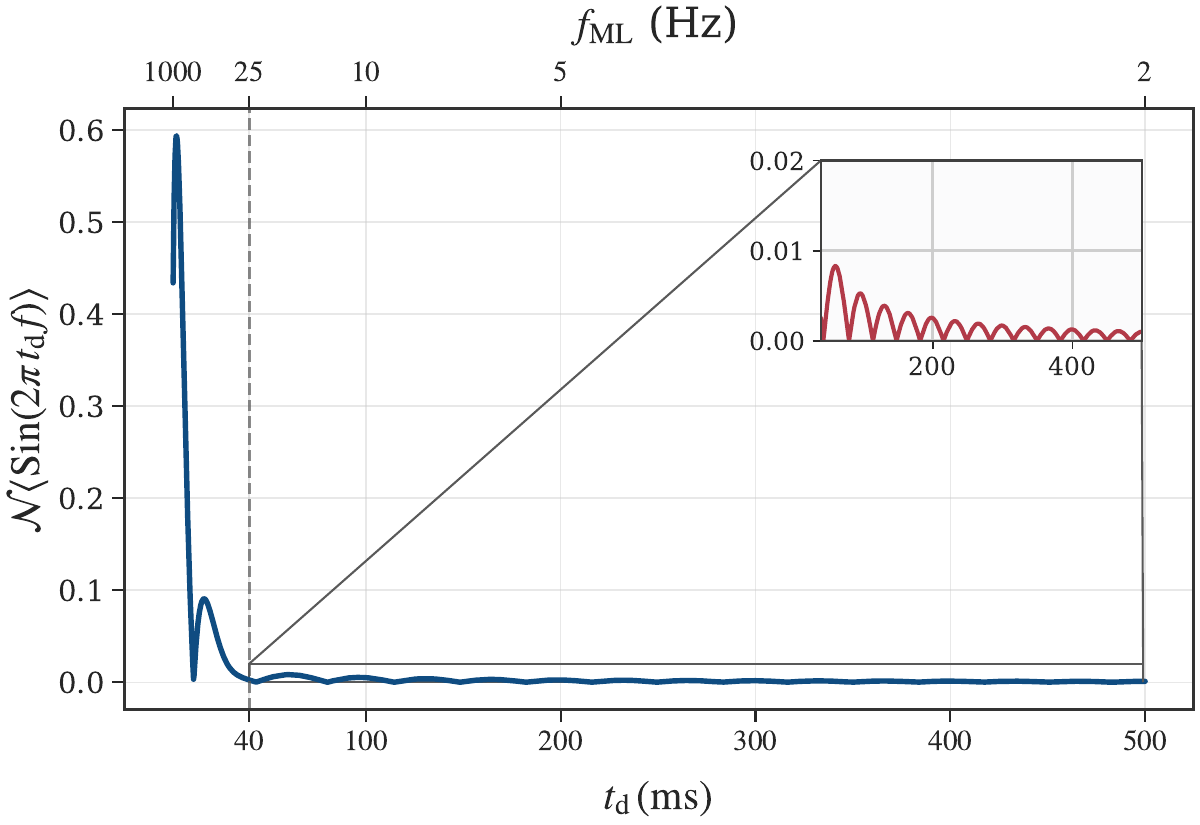}
    \caption{The variation of the average of $\mathcal{N}\sin{(2\pi f t_{\mathrm{d}})}$ evaluated as a function of $t_{\mathrm{d}}$. The average is less than $1/100$ for $t_{\mathrm{d}} \geq 40$ ms, which can be treated as practically zero under the geometric-optics approximation.}
\label{fig:sine_term}
\end{figure}

From Fig.~\ref{fig:mismatch_plot_TF2}, we observe a boundary of the low-$FF$ region, which becomes independent of $M_{\mathrm{Lz}}$ for a given $y$. This boundary parallel to $M_{\mathrm{Lz}}$-axis is roughly the same across signals of different $M_{\mathrm{tot}}$. This behavior can be understood when one considers the geometric-optics approximation to the lensing amplification factor. 
Consider the match between microlensed and unlensed waveforms given by Eq.~(\ref{eq:match_ml_ul}), which we evaluate explicitly in the geometric-optics regime.
We define the average value of a quantity $Q(f)$, say $\langle Q \rangle$ in the following way: the average is computed with respect to a weight function which is the signal power divided by the PSD, assuming an O4-like (design sensitivity), of the noise, 
\begin{equation}
\langle Q \rangle \equiv 
\frac{1}{\mathcal N}
\int_{f_{\mathrm{low}}}^{f_{\mathrm{high}}}
\frac{f^{-7/3}\, Q(f)}{S_{\mathrm{h}}(f)}\,df ,
\end{equation}
where the normalisation factor $\mathcal{N}$ is defined as,
\begin{equation}
\mathcal{N} = 
\int_{f_{\mathrm{low}}}^{f_{\mathrm{high}}}
\frac{f^{-7/3}}{S_{\mathrm{h}}(f)}\,df.
\end{equation}
We can write the unlensed and lensed waveform as,
\begin{align}
    h_{\rm UL} (f;\lambda_{\mathrm{CBC}}) &= A(\lambda_{\mathrm{CBC}})f^{-7/6} \, e^{i\Phi_{\mathrm{CBC}} (f;\lambda_{\mathrm{CBC}})} ,\\ \nonumber
    h_{\rm L} (f;\lambda_{\mathrm{L}}, \lambda_{\mathrm{CBC}}) &= |F(f; \lambda_{\mathrm{L}})| ~ (A(\lambda_{\mathrm{CBC}}) f^{-7/6}) \nonumber\\ 
        &\quad \times e^{i \left(\Phi_{\mathrm{L}}(f; \lambda_{\mathrm{L}}) + \Phi_{\mathrm{CBC}}(f; \lambda_{\mathrm{CBC}}) \right)}\label{eq:lensed_polar}.
\end{align}
Thus, it follows from the above equations that,
\begin{align}
    \left( h_{\mathrm{UL}} | h_{\mathrm{UL}}  \right) &= \int_{f_{\mathrm{min}}}^{f_{\mathrm{max}}} \frac{| h_{\mathrm{UL}}(f)|^2}{S_{\mathrm{h}}(f)} df \nonumber \\
    &= A^2 \mathcal{N}
\end{align}
and,
\begin{align}
    \left( h_{\mathrm{L}} | h_{\mathrm{L}}  \right) &= \int_{f_{\mathrm{min}}}^{f_{\mathrm{max}}} \frac{| h_{\mathrm{L}}(f)|^2}{S_{\mathrm{h}}(f)} df \nonumber \\&= \int_{f_{\mathrm{min}}}^{f_{\mathrm{max}}} \frac{A^2 f^{-7/3}| F(f)|^2}{S_{\mathrm{h}}(f)} df. 
\end{align}
It can be shown that under the geometric optics approximation,
\begin{align}
\int_{f_{\mathrm{min}}}^{f_{\mathrm{max}}} \frac{| h_{\mathrm{L}}(f)|^2}{S_{\mathrm{h}}(f)} \, df 
&= A^2\mathcal{N} \Big[ \langle |\mu_{+}| + |\mu_{-}| \rangle \nonumber \\
&\quad + 2 \sqrt{|\mu_{+}||\mu_{-}|} \langle \sin(2\pi f t_{\mathrm{d}}) \rangle \Big]
\end{align}
From Fig.~\ref{fig:sine_term}, it can be seen that for $t_{\mathrm{d}} \geq 40$ ms, the average of the second oscillating term is small. In the most extreme case of $y \to 0$, where $|\mu_{+}| \approx |\mu_{-}|$, the sine term is 100 times smaller than the first term, thus,
\begin{align}\label{eq:geom_sig_norm_approx}
\int_{f_{\mathrm{min}}}^{f_{\mathrm{max}}} \frac{| h_{\mathrm{L}}(f)|^2}{S_{\mathrm{h}}(f)} \, df  \approx A^2\left(|\mu_+| + |\mu_-|\right).
\end{align}
The match $\mathcal{M} = (h_{\mathrm{L}} | h_{\mathrm{UL}})_{\mathrm{max}\{t_{\mathrm{c}}, \phi_{\mathrm{c}}\}}$ can be computed as, 
\begin{align}
\mathcal{M} 
&\approx \frac{1}{ \mathcal{N} \sqrt{(|\mu_+| + |\mu_-|)}} \nonumber \\
&\quad \times \Bigg[
\int_{f_{\mathrm{low}}}^{f_{\mathrm{high}}}
\frac{f^{-7/3}\, \sqrt{|\mu_{+}|} \, \mathrm{e}^{2\pi i f t_{\mathrm{c}}}}{S_{\mathrm{h}}(f)}\, df \nonumber \\
&\quad -i \int_{f_{\mathrm{low}}}^{f_{\mathrm{high}}}
\frac{f^{-7/3}\, \sqrt{|\mu_{-}|} \, \mathrm{e}^{2\pi i f (t_{\mathrm{c}} + t_{\mathrm{d}} )}}{S_{\mathrm{h}}(f)}\, df 
\Bigg].
\end{align}
As $\sqrt{|\mu_+}| > \sqrt{|\mu_-}|\quad \forall\quad y > 0$, the above quantity is maximized for $t_{\mathrm{c}} = 0$, i.e, when the unlensed template is aligned with the image having higher magnification of $\sqrt{|\mu_+}|$. Thus, the corresponding match obtained is independent of $M_{\mathrm{Lz}}$, and is approximately given by,
\begin{equation}
    \mathcal{M} \approx \frac{\sqrt{|\mu_+}|}{\sqrt{|\mu_+| + |\mu_-|}}.
\end{equation}

\section{Constructing the lensing template bank}
\label{sec:method}

The template placement in the full parameter space, consisting of the CBC and the lens parameters, can be obtained from a metric in principle. The form of the metric for this space has been obtained in Appendix \ref{appendix:phase_metric}. However, we do not expect much cross talk between $\lambda_{\mathrm{CBC}}$ and  $\lambda_{\mathrm{L}}$. Our numerical computations strongly support this expectation. We tested this by considering a Newtonian chirp signal characterized by single parameter $\tau_0$. Thus, the full metric $g_{ab}$ is a $3\times3$ matrix, whose proper volume is defined as the square root of the absolute value of the determinant of the metric, $\sqrt{|\det{g}|}$ integrated over the full parameter space. We evaluate this integral corresponding to the exact case, and the other by taking the block diagonal form by zeroing out the matrix elements outside the blocks, we find that difference in volumes is less than $0.1$ \%, for our range of parameters. Therefore, the metric is almost exactly block diagonal with two blocks one pertaining to CBC and the other to the lens; the full metric can be written as as a sum of the two pieces:
\be
g_{a b} \Delta \Lambda^a \Delta \Lambda^b = g_{\alpha \beta} \Delta \lambda^\alpha \Delta \lambda^\beta ~+~ g_{\ap \bp} \Delta \lambda'^\ap \Delta \lambda'^\bp \,,
\ee
where $\a, \b$ run over the CBC parameters and $\ap, \bp$ run over the lensing parameters and $a = \{\a, \ap \}, b = \{\b, \bp\}$ are the full set of parameters. In effect, we have a {\it product} template bank, which is a Cartesian product of the CBC bank and the lensing bank. This simplification greatly facilitates the computations while at the same time also elucidating our understanding.

The Cartesian product structure of the product template bank reduces the number of templates that need to be generated. Indeed, were it not for this structure, every point in the unlensed CBC parameter space spanned by the corresponding bank would require a new lensing template bank, with template choice and placement in the lensing template bank becoming a function of the CBC parameters. This not being the case, we only require to construct the lensing template bank once, and can readily ``tack it on'' to any set of unlensed CBC templates. In the subsection below, we present the computation of the lensing metric and the corresponding lensing template bank.

\subsection{The phase metric for the lens parameters}

On similar lines of the discussions in the Appendix \ref{appendix:phase_metric}, we derive the expression for the phase matrix for $\Phi_{\mathrm{L}}$. We use the subscript $\mathrm{L}$ and $\mathrm{CBC}$ to denote the lensing and CBC quantities, respectively, instead of the prime and unprimed quantities used in the Appendix. Consider the lensed signal $h_{\mathrm{L}}$ given by Eq.~(\ref{eq:lensed_polar}), which is normalized as follows,
\begin{equation}
    \hat{h}_{\mathrm{L}} = \frac{h_{\mathrm{L}}}{|h_{\mathrm{L}}|} = \displaystyle \frac{|F(f)| f^{-7/6} e^{i (\Phi_{\mathrm{L}} + \Phi_{\mathrm{CBC}})}}{\left[ \int_{f_{\mathrm{low}}}^{f_\mathrm{high}} \frac{f^{-7/3} |F(f)|^2}{S_{\mathrm{h}}(f)} ~ df \right]^{1/2}}.
\end{equation}
where $\,\hat{}\,$ denotes that the signal has been normalized. Assuming geometric optics approximation for $F(f)$, the amplitude and phase can be written as,
\begin{align}\label{eq:mod_Ff}
    |F_{\mathrm{GO}}(f)| &= \left[ \mathrm{Re}(F_{\mathrm{GO}})^2 + \mathrm{Im}(F_{\mathrm{GO}})^2 \right]^{1/2} \nonumber \\
    &= \sqrt{|\mu_-|} \left[ \left( \mu_{\mathrm{r}} \right)^2 + 1 + 2\mu_{\mathrm{r}} \sin{(2\pi f t_d)} \right]^{1/2},
\end{align}
and,
\begin{align}\label{eq:arg_Ff}
    \Phi_{\mathrm{L}}(f) &= \tan^{-1} \left[ \frac{\mathrm{Im}(F_{\mathrm{GO}})}{\mathrm{Re}(F_{\mathrm{GO}})}\right] \nonumber \\
    &= \tan^{-1}\left[ \frac{-\cos{(2\pi f t_{\mathrm{d}})}}{\mu_{\mathrm{r}} + \sin{(2 \pi f t_{\mathrm{d}})}}\right],
\end{align}
where $\mu_{\mathrm{r}} \equiv \sqrt{|\mu_{+}|} / \sqrt{|\mu_{-}|}$. From Eq.~(\ref{eq:geom_sig_norm_approx}), the normalized lensed signal takes the form
\begin{align}
    \hat{h}_{\mathrm{L}} \approx  \left[ 1 + \frac{\mu_{\mathrm{r}}}{1 + \mu_{\mathrm{r}}^2}
    \sin{(2\pi f t_{\mathrm{d}})} \right] 
    \frac{f^{-7/6} e^{i (\Phi_{\mathrm{L}} + \Phi_{\mathrm{CBC}})}}{\sqrt{\mathcal{N}}}
\end{align}   
Perturbing only the lens parameters, the match $\mathcal{M}$ between signals with lens parameters $\lambda_{\mathrm{L}}$ and $\lambda_{\mathrm{L}} + 
\Delta\lambda_{\mathrm{L}}$ is
\begin{equation}
    \mathcal{M}(\lambda_{\mathrm{L}}, \Delta\lambda_{\mathrm{L}}) =
    \bigl(\, \hat{h}_{\mathrm{L}}(f;\lambda_{\mathrm{L}}) 
    \mid 
    \hat{h}_{\mathrm{L}}(f;\lambda_{\mathrm{L}} + \Delta\lambda_{\mathrm{L}}) \,\bigr).
\end{equation}
The amplitude prefactor of $\hat{h}_{\mathrm{L}}$ inside the square brackets is an oscillating function of frequency. Since the match is weakly sensitive to the amplitude profile, we replace it by its average value. The match integral then reduces to,
\begin{align}
    \mathcal{M}(\lambda_{\mathrm{L}}, \Delta\lambda_{\mathrm{L}}) &=
    \frac{1}{\mathcal{N}}
    \left|
    \int_{f_{\mathrm{low}}}^{f_{\mathrm{high}}}
    \frac{
        e^{i\left[
            \Phi_{\mathrm{L}}(f;\lambda_{\mathrm{L}}+\Delta\lambda_{\mathrm{L}})
            -
            \Phi_{\mathrm{L}}(f;\lambda_{\mathrm{L}})
          \right]}
      }{f^{7/3}\,S_{\mathrm{h}}(f)}
    \, df
    \right|.
\end{align}
\begin{figure*}[t]
    \centering
    \includegraphics[width=\textwidth]{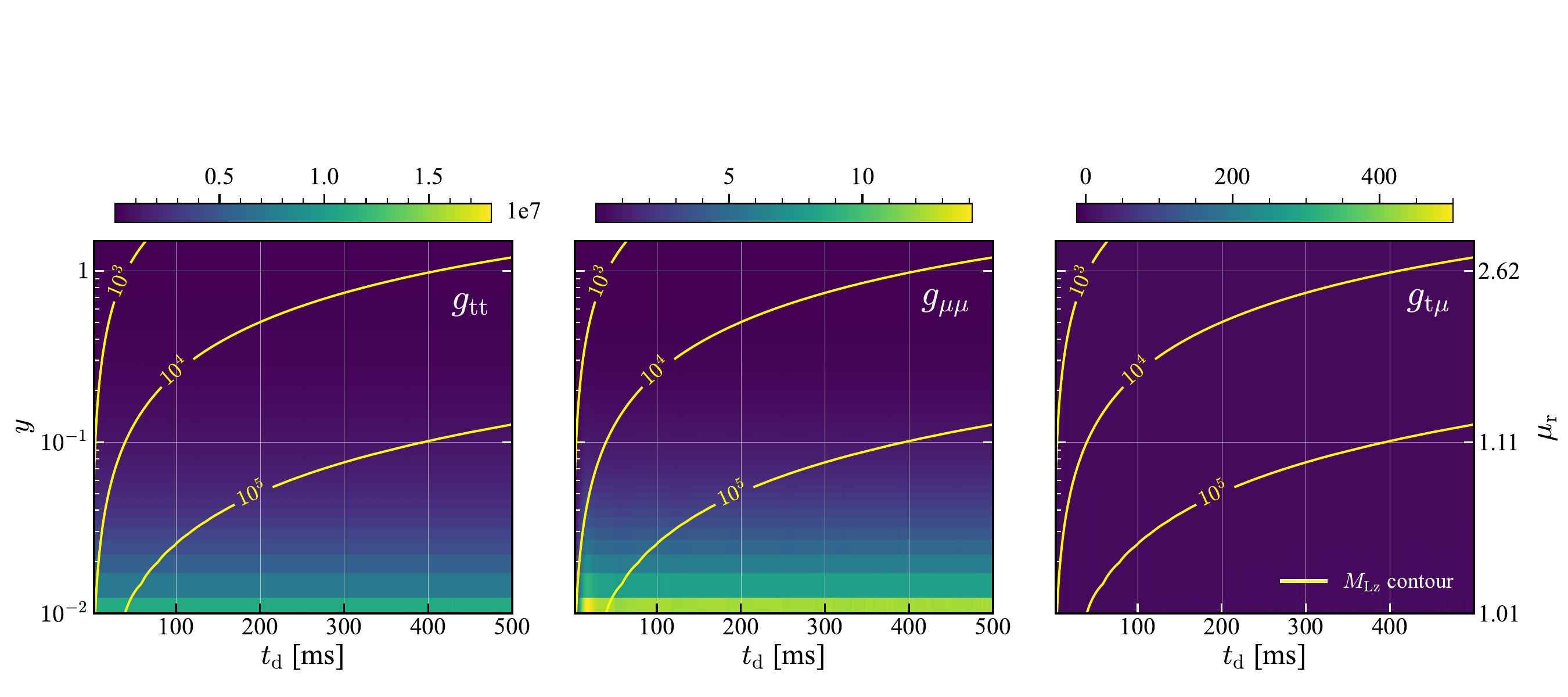}
    \caption{The metric elements $g_{\mathrm{tt}}$(left), $g_{\mu \mu}$ (middle) and $g_{\mathrm{t \mu}}$ (right) shown as a function of $t_{\mathrm{d}}-\mu_{\mathrm{r}}$. Although we computed metric elements over a square grid in the range $t_{\mathrm{d}} \in [1-500]$ ms and $\mu_{\mathrm{r}}\in (1, 5.5]$, We are interested in only those parameters having $10^2 \leq M_{\mathrm{Lz}} \leq 10^5 M_{\odot}$, shown by yellow contours.}
\label{fig:metric_elements}    
\end{figure*}
\begin{equation}
    \mathcal{M}(\lambda_{\mathrm{L}}, \Delta\lambda_{\mathrm{L}}) = 1 - (g_{\mathrm{L}})_{\alpha \beta} \Delta \lambda_{\mathrm{L}}^{\alpha} \Delta \lambda_{\mathrm{L}}^{\beta}
\end{equation}
and that the resulting metric is given by,
\begin{equation}
\label{eq:lensing_metric}
    (g_{\mathrm{L}})_{\alpha \beta} = \frac{1}{2} \left( \langle (\Phi_{\mathrm{L}})_{\alpha} (\Phi_{\mathrm{L}})_{\beta} \rangle - \langle (\Phi_{\mathrm{L}})_{\alpha} \rangle \langle (\Phi_{\mathrm{L}})_{\beta} \rangle \right) \,.
\end{equation}
Here, $(\Phi_{\mathrm{L}})_{\alpha} \equiv \frac{\partial \Phi_{\mathrm{L}}}{\partial \lambda^{\alpha}}$. The metric is, essentially, the covariance matrix.
\par
\subsection{Lensing template bank construction}
For the construction of a geometric template bank, it is desirable to choose coordinates such that the phase has a linear dependence on them. Such a choice of coordinates makes the metric constant throughout the parameter space. However, it may not always be possible to find such a transformation. Neverthless, from Eq.~(\ref{eq:arg_Ff}), we choose $\lambda_{\mathrm{L}} \in \{ t_{\mathrm{d}}, \mu_{\mathrm{r}} \}$ as the parameters for the  lensing template bank construction. The derivative of $\Phi_{\mathrm{L}}$ with respect to the lensing parameters is given by,
\begin{figure}[]
    \centering
    \includegraphics[width=\columnwidth]{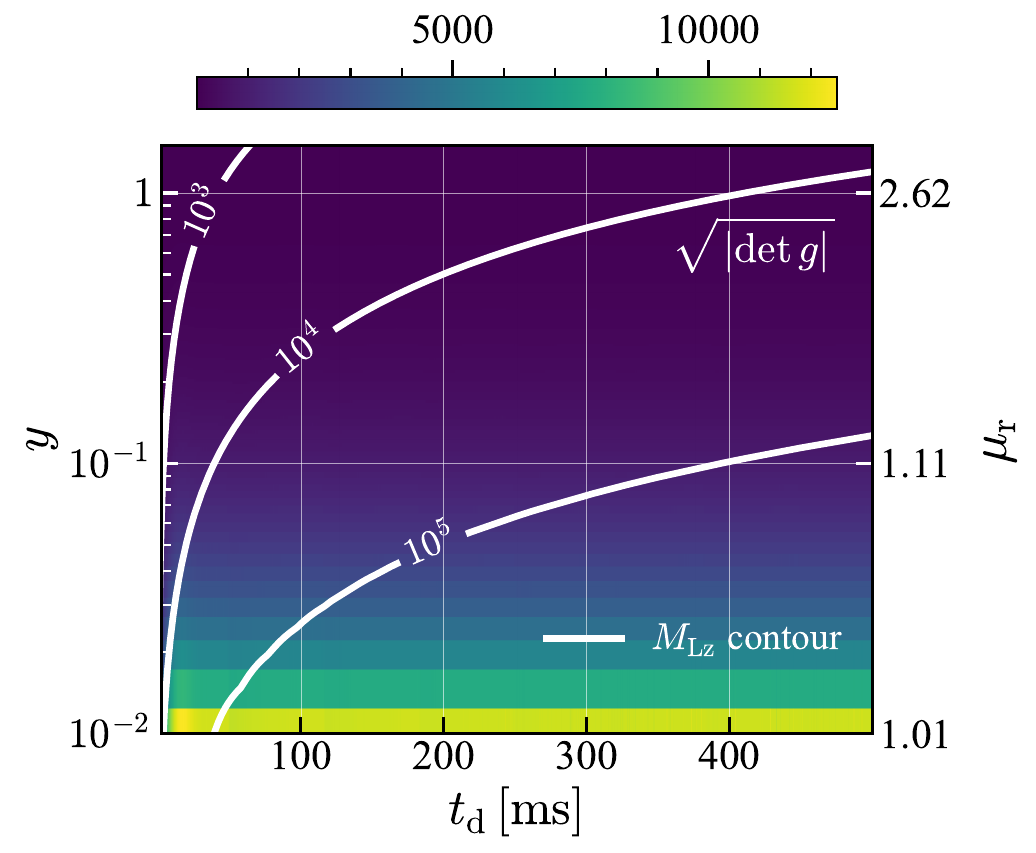}
    \caption{$\sqrt{|\mathrm{det}(g)}|$ as function of $t_{\mathrm{d}} - \mu_{\mathrm{r}}$. The proper volume of this surface corresponding to parameters $10^2 \leq M_{\mathrm{Lz}} \leq 10^5 M_{\odot}$ is evaluated numerically and it turns out to be 135. The proper volume occupied by a single template is $2\epsilon = 0.06$. Thus, the minimum number of templates turn out to be around 2250.}
    \label{fig:sqrt_det_metric}
\end{figure}

\begin{subequations}
   \begin{align}
(\Phi_{\mathrm{L}})_{t_{\mathrm{d}}}(f)
&=
\frac{2\pi f \,\bigl( 1 + \mu_{\mathrm{r}} \sin(2\pi f t_{\mathrm{d}}) \bigr)}
     {1 + (\mu_{\mathrm{r}})^{2} + 2\mu_{\mathrm{r}} \sin(2\pi f t_{\mathrm{d}})},
\\[6pt]
(\Phi_{\mathrm{L}})_{\mu_{\mathrm{r}}}(f)
&=
\frac{\cos(2\pi f t_{\mathrm{d}})}
     {1 + (\mu_{\mathrm{r}})^{2} + 2\mu_{\mathrm{r}} \sin(2\pi f t_{\mathrm{d}})}.
\end{align} 
\end{subequations}
The phase derivatives are dependent on the coordinates, hence the lensing metric $g_{\mathrm{L}}$ in these coordinates will not be constant across the parameter space. From the $FF$ plots in Fig. \ref{fig:mismatch_plot_TF2}, we identify the region of parameter space where the lensing template bank is required. The region of interest, where $FF$ values fall below $97\%$ is roughly characterized by  
\begin{itemize}
    \item $0.01 \leq y \leq 2$ which corresponds to $(1 < \mu_{\mathrm{r}} \leq 5.5)$, and

    \item $2 \leq f_{\mathrm{ML}} \leq 1000$ Hz, i.e, $1 \leq t_{\mathrm{d}} \leq 500$ ms.

    \item $10^2 \leq M_{\mathrm{Lz}} \leq 10^5~M_{\odot}$
\end{itemize}
It has to be noted that although the low-$FF$ region extends beyond $t_{\mathrm{d}} = 500$ ms (or, $f_{\mathrm{ML}} < 2$ Hz), we choose not to lay templates in the far geometric-optics regime as lensed signals with $t_{\mathrm{d}} > 500$ ms would have SNR peaks well resolved in time. Consequently, the lensed CBC may be detected as temporally separated signals in the standard CBC search pipelines. 

We compute the lensing metric numerically from Eq. (\ref{eq:lensing_metric}), whose elements are shown in Fig. \ref{fig:metric_elements}. Also, Fig. \ref{fig:sqrt_det_metric} shows the variation of the square root of the determinant of the metric across the parameter space. The expected number of lensing templates $N_{\mathrm{L}}$ can be estimated as the ratio of the proper volume of the parameter space to the proper volume of a single template as,
\begin{equation}\label{eq:template_number_estimate}
    N_{\mathrm{L}} = \frac{\int dt_{\mathrm{d}} d \mu_{\mathrm{r}} \sqrt{| \mathrm{det}(g_{\mathrm{L}}) |}}{dl^2},
\end{equation}
\begin{figure}[]\label{fig:mismatch_square}
    \centering
    \includegraphics[width=\columnwidth]{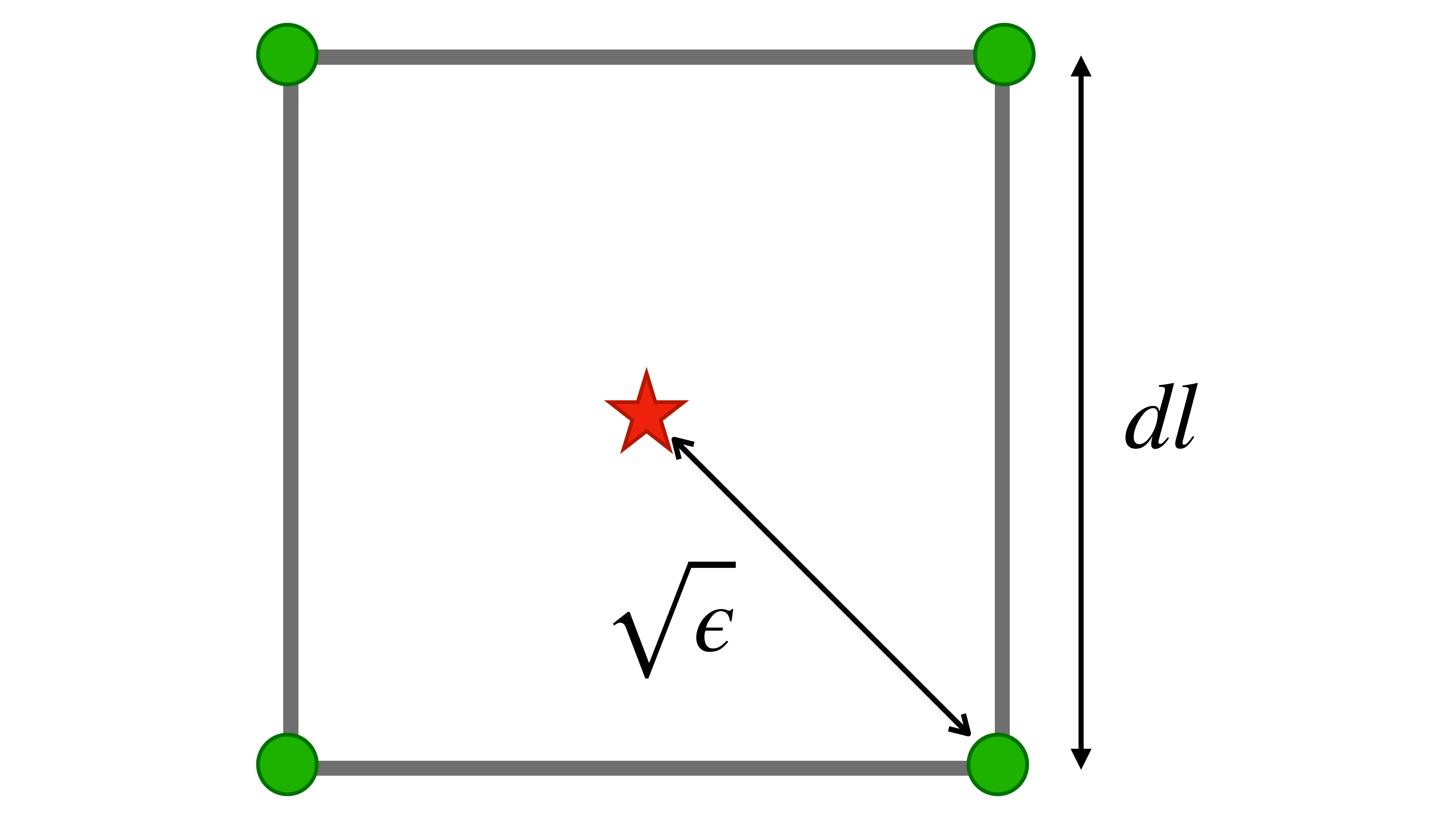}
    \caption{An illustration of the Mismatch Square. The green circles represent the template placement and the red star for the signal parameter. It can be seen that for 2D configuration, $dl^2 = 2\epsilon$.\footnote{A square placement of templates is likely not optimal. Indeed, given a flat metric, a hexagonal placement is more optimal. Nevertheless, we don't expect the number of templates required to change appreciably with a more optimal placement strategy. Moreover, the performance of the template bank with square placement is found to be more than adequate at improving the search sensitivity, as shown in Section~\ref{sec:results}.}}
    \label{fig:mm_square}
\end{figure}
where ($dl$) is the distance between two nearest templates specified by the minimum match criterion. We require $dl$ to be such that the maximum allowed mismatch between signal and nearest template should be $\epsilon = 0.03$. This can be computed as follows: Consider the templates arranged at the corners of a square of side $dl$, as depicted in Fig.~\ref{fig:mm_square}. The worst possible case is considered to be when the point corresponding to the signal is exactly in the center of the square. We, thus require that the distance between the signal and any of the vertices of square to be $\sqrt{\epsilon}$. From Fig.~\ref{fig:mm_square}, it can be seen that $dl = \sqrt{2\epsilon}$, which we use in Eq.~(\ref{eq:lensing_metric}) by setting $ds^2 = 2\epsilon$ for template placement.

We evaluate the numerator of Eq.~(\ref{eq:template_number_estimate}) numerically, limiting ourselves to only those $t_{\mathrm{d}}-\mu_{\mathrm{r}}$ parameters satisfying $10^2 \leq M_{\mathrm{Lz}} \leq 10^5 M_{\odot}$, and find it to be 135. For $\epsilon = 0.03$, the optimum number of templates turns out to be 2250 for the selected parameter space. To simplify the template placement, we ignore the off-diagonal term $(g_{\mathrm{L}})_{\mathrm{t \mu}}$. This step is justified as we find that the numerical evaluation of the proper volume of the parameter space, when evaluated without considering the off-diagonal terms, also turns out to be approximately 135, indicating negligible correlation between the lensing parameters. 
%

\begin{figure*}[h!t]
    \centering
    \includegraphics[width=\textwidth]{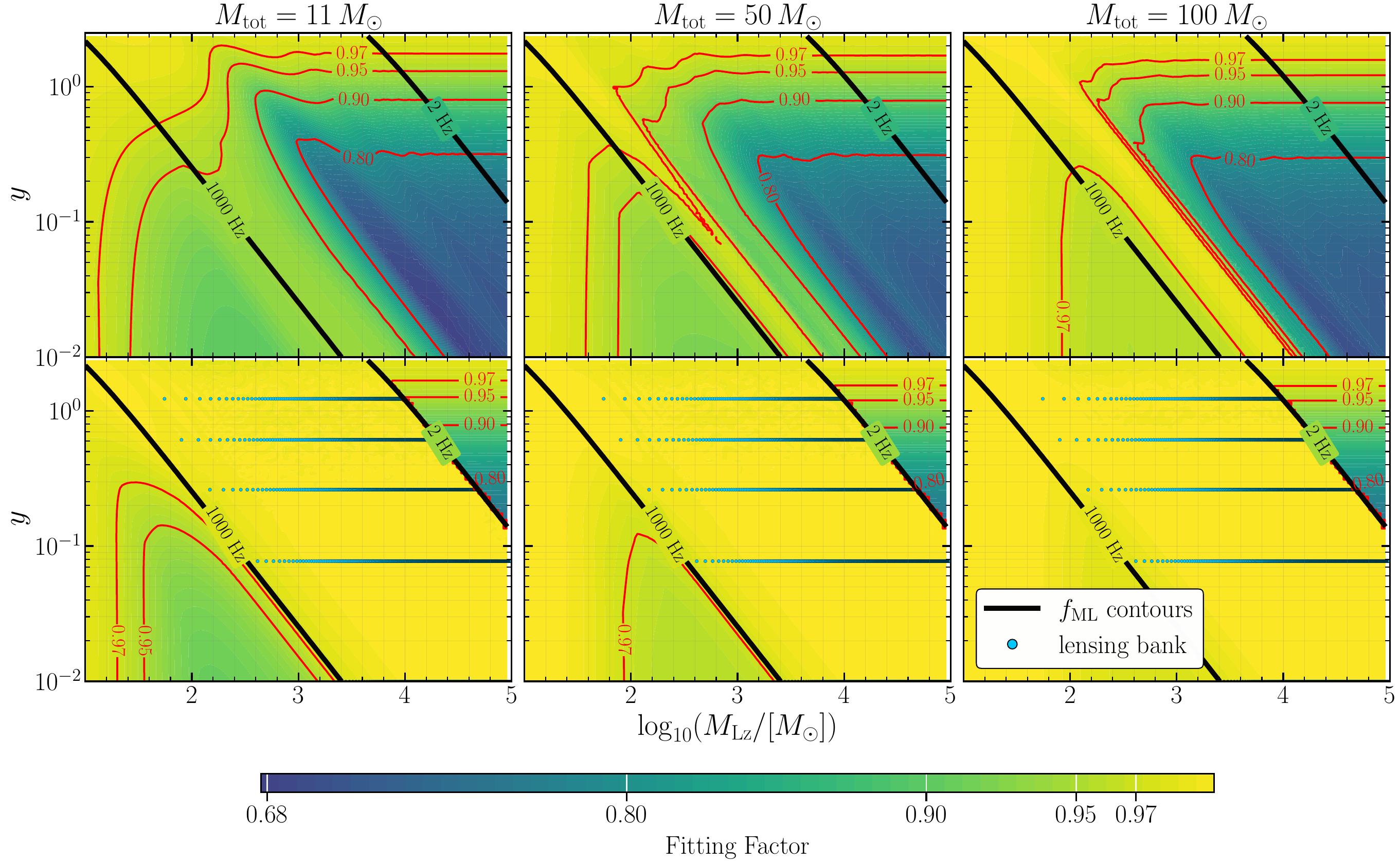}
    \caption{$FF$ computed for microlensed signals and the unlensed templates (top panel), computed for $M_{\mathrm{Lz}} \in \{ 11, 50, 100 \}\, M_{\odot}$ arranged from left to right. Improved $FF$ plots are shown in the bottom panel after the inclusion of the product template bank in the fitting-factor analysis. Lensing bank parameters are shown as blue circles.}
\label{fig:Improved_FittingFactor}    
\end{figure*}
From Fig.~\ref{fig:metric_elements}, it can be seen that the metric element $(g_{\mathrm{L}})_{\mathrm{tt}}$ is roughly constant for $t_{\mathrm{d}} \geq 20 $ ms. As a consequence, the template density in this region remains roughly constant in the $t_{\mathrm{d}}$ direction. This region also corresponds to the region having parameters with $M_{\mathrm{Lz}} \geq 10^3 M_{\odot}$, where the template bank is needed the most. From these observations, we place the templates in the $t_{\mathrm{d}}-\mu_{\mathrm{r}}$ space as follows.
\begin{itemize}
    \item The metric used for template placement is given by

    \begin{equation}\label{eq:lensing_metric_diagonal}
        \Delta s^2 = (g_{\mathrm{L}})_{\mathrm{tt}} (\Delta t_{\mathrm{d}})^2 + (g_{\mathrm{L}})_{\mathrm{\mu \mu}} (\Delta \mu_{\mathrm{r}})^2 = 2\epsilon
    \end{equation}

    \item Begin with template placement in the $\mu_{\mathrm{r}}$ direction for a fixed $t_{\mathrm{d}}$. From the Eq.~(\ref{eq:lensing_metric_diagonal}), the template spacing in the $\mu_{\mathrm{r}}$ direction turn out to be, $\Delta \mu_{\mathrm{r}} = \sqrt{\frac{2 \epsilon}{(g_{\mathrm{L}})_{\mathrm{\mu \mu}}}}$. 

    \item Having placed the templates in the $\mu_{\mathrm{r}}$ direction for a given $t_{\mathrm{d}}$, we update the next point in $t_{\mathrm{d}}$ direction according to $\Delta t_{\mathrm{d}} = \sqrt{\frac{2 \epsilon}{\overline{(g_{\mathrm{L}})}_{\mathrm{t t}}}}$, where $\overline{(g_{\mathrm{L}})}_{\mathrm{t t}}$ is the average value of the metric element for a given $t_{\mathrm{d}}$. We successively place the templates until the entire parameter space is covered.
\end{itemize}
Using this strategy, we get around 4032 templates. The bottom panel of Fig. \ref{fig:Improved_FittingFactor} shows the template bank parameters placement in the $M_{\mathrm{Lz}}-y$ space.

\section{Demonstration of the product template bank}\label{sec:results}

We validate the efficacy of the product template bank by computing the fitting-factor for microlensed signals using the product template bank for a fixed CBC system. The top panel of Fig. \ref{fig:Improved_FittingFactor} shows the Fitting factor plots when computed for CBC systems for total mass of $\{11, 50, 100\} M_{\odot}$ and $q=1$, from left to right, when computed with respect to a non-spinning unlensed template bank. The CBC templates were generated using \textsc{IMRPhenomD} waveform model. 

Consistent with the previous trends observed in Sec~\ref {sec:motivation}, we observe that the low fitting-factor region shrinks as the CBC becomes heavier. The bottom panel shows the fitting factor plot after the product template bank is included in the analysis. We also show the placement of the lensing templates (shown as blue circles) on top of the fitting factor plot in the bottom panel. We find that the fitting factor improves within the band $f_{\mathrm{ML}} \in [2, 1000]$ Hz, consistent with the prescribed minimum match criteria of $0.97$ in the region where templates are placed. Since the available sensitive volume scales as $\sim \mathcal{M}$, we observe a significant improvement in the search sensitive volume, up from $(0.7)^3 \sim 35\%$ to $(0.97)^3 \sim 91\%$. We don't place templates below $f_{\mathrm{ML}} =  2$ Hz, because this corresponds to overlapping signals with $t_{\mathrm{d}} \geq 0.5$ seconds. As mentioned earlier, such lensed signals may be detected as two distinct signals in the search pipelines, as their SNR peaks will be well separated in time. 

For lighter systems, we notice an additional region of low fitting factor. As pointed out in Fig.~\ref{fig:mismatch_plot_IMR}, this effect is not captured by the geometric-optics approximation. But we also notice that the loss of SNR in this region is not as severe as the loss occurring due to overlapping signals. Moreover, these regions span mostly low impact parameters, which have low probability of occurrence and disappear successively for heavier systems. Since heavier systems have a larger probability of lensing due to larger detection horizon and thus, greater optical depth, this region is not much of our concern. Thus, for most practical applications, the lensing bank serves its purpose. 

\section{Summary and Outlook}\label{sec:summary_outlook}
The identification of microlensing signatures in CBC signals is a high profile target of GW astronomy. Templated searches typically use quasi-circular templates, and are therefore not optimally tuned to target microlensed signals, resulting in considerable loss of sensitivity to such signals. 

In this work, we present for the first time, a prescription to construct a product geometric template bank for microlensed GW signals from CBCs, assuming a point-mass lens model. We first provide a fitting factor study assessing the recovery of microlensed signals with unlensed templates. We find that the matches between the two waveforms, maximized over the unlensed CBC signal parameter space, results in a significant loss of sensitivity for a large portion of the lens parameter space considered, thereby motivating the need for a lensing template bank.

To construct the lensing template bank, we leveraged the analytic form of the lensing phase $\Phi_{\mathrm{L}}(f)$ in the geometric-optics regime and derived the corresponding phase metric in the transformed $ t_{\mathrm{d}} - \mu_{\mathrm{r}}$ parameter space. We also argued that the full parameter-space metric is nearly block-diagonal, with negligible cross-talk between the CBC and lensing parameters. 
 
The full product template bank, thus can be obtained as a Cartesian product of this lensing bank with the existing unlensed CBC template bank. Although we have done this study for non-spinning CBC signals, the derivation of the lensing phase matrix does not assume any specific form of the unlensed signal. Therefore, the lensing bank can be tacked on to the unlensed CBC template banks with additional physical content such as spins and eccentricity.

We validated the lensing bank's efficacy by evaluating the fitting factor across the lens parameter space for fiducial values of the unlensed CBC parameters. The inclusion of the lensing bank recovers the microlensed signals -- injected using the full analytic amplification factor -- with the prescribed minimal match criterion of $0.97$ or larger, thereby demonstrating that the geometric-optics approximation captures the essential features of microlensing relevant for template placement. 

Matched filtering a product template bank could be computationally intensive because the size of the unlensed template bank itself is large, consisting of $\mathcal{O}(10^5-10^6)$ templates. Nevertheless, the lack of correlations between lens and CBC parameters can be exploited to argue that while there is non-trivial loss of SNR when microlensed CBCs are recovered with unlensed templates, the best matched unlensed templates are expected to lie in the vicinity of the true values of the microlensed CBC's intrinsic parameters. A hierarchical search strategy can therefore be employed, where product template banks targeting superthreshold CBCs, are constructed as Cartesian products between the templates spanning the neighborhood of the best matched CBC template, and the lensing template bank. Hierarchical search strategies have been devised and explored in the literature \cite{mohantysvd96,mohantysvd98,anadsensvd2002,anandsensvd2003,bhooshan_hierarchical,kanchan_hierarchical} before and we expect that they would be useful in this context. We leave a detailed exploration of these strategies, in the context of false-alarm rates and detection statistics, to future work.

The product template bank can be used in a preliminary search for microlensing signatures in CBCs observed marginally below the ranking statistic (SNR) threshold. Typically, such CBCs are not further investigated to search for microlensing signatures, because of the prohibitive computational costs of large scale parameter estimation runs. Constructing product template banks targeting such CBCs could enable a rapid, and computationally inexpensive means, of determining whether the subthreshold CBC candidate is pushed above threshold upon inclusion of microlensing effects in the templates. This provides an efficient method to either rule out subthreshold CBCs as not microlensed, or mark them up as sufficiently interesting for further microlensing investigations. 

This work would also be useful in the context of future third-generation detectors such as the Cosmic Explorer \cite{CE} and the Einstein Telescope \cite{ET}. Because of enhanced sensitivity, the number of merger events is expected to increase drastically, by $10^3$ or more. This, in turn, would increase the chance of observing microlensed events. A product template bank such as the one we describe in this work, in tandem with a hierarchical search strategy mentioned above, will be critical in helping to identify microlensed events in a computationally feasible way. 

\textit{Software}: \texttt{PyCBC} \citep{pycbc}, \texttt{GWMAT} \citep{gwmat}, \texttt{NumPy} \citep{vanderWalt:2011bqk}, \texttt{SciPy} \citep{Virtanen:2019joe}, \texttt{astropy} \citep{2013A&A...558A..33A, 2018AJ....156..123A}, \texttt{Matplotlib} \citep{Hunter:2007}, \texttt{jupyter} \citep{jupyter}.

\section*{Acknowledgements}
We thank Anand S. Sengupta for reading the
manuscript and comments. SG thanks Anirban Kopty and Kanchan Soni for useful discussions. SG's research was supported by the University Grants Commission, Government of India. SJK gratefully acknowledges support from the Science and Engineering Research Board (SERB) through Grant No. SRG/2023/000419. We acknowledge the use of IUCAA LDG cluster Sarathi for the computational work. We are grateful for computational resources provided by the Leonard E. Parker Center for Gravitation, Cosmology, and Astrophysics at the University of Wisconsin-Milwaukee. 
 
\section*{Data Availibility}
The data that support the findings of this article are not
publicly available. The data are available from the authors
upon reasonable request.

\section{Appendices}\label{sec:appendix}

\subsection{Lensing Amplification factor for point-mass lens model}\label{subsec:microlens_ff_derivation}

 In the lens-plane, we choose the coordinate axes such that $\mathbf{y} = (y, 0)$ and $\mathbf{x} = (r\cos (\theta), r \sin (\theta))$. Thus, we can write,

\begin{equation}
    \left| \mathbf{x} - \mathbf{y}\right|^2 = r^2 + y^2 -2ry \cos(\theta) \,.
\end{equation}

The lensing potential and $\phi_{\mathrm{min}}$ for the point -mass lens model are given by,

\begin{align}
    \psi_{\mathrm{L}}(\mathbf{x}) &= \ln (r) \,, \\
    \phi_{\mathrm{min}}(\mathbf{y}) &= \frac{(x_{\mathrm{m}} - y)^2}{2} - \mathrm{ln}(x_{\mathrm{m}}) \,, 
\end{align}
where $x_{\mathrm{m}}= \frac{y + \sqrt{y^2 + 4}}{2}$. Therefore, the diffraction integral in Eq. (\ref{eq:diffraction_integral}) can be written as,

\begin{align}
    &F(\omega, y) 
    = \frac{\omega}{2\pi i}
    \exp\!\left[i\omega\!\left(\frac{y^2}{2}-\phi_{\mathrm{min}}(y)\right)\right] \nonumber \\
    &\quad \times \int_{0}^{2\pi} d\theta \int_{0}^{\infty} r\,dr \,
    \exp\!\left\{ i\omega
    \left[
    \frac{r^2}{2}
    - r y \cos(\theta) - \ln(r)
    \right]
    \right\}.
\end{align}
The angular integral is related to the zeroth-order Bessel function as,

\begin{equation}
\int_{0}^{2\pi} e^{\,i \omega r y \cos{\theta}} d\theta
= 2\pi J_0(\omega r y).
\end{equation}
We therefore obtain,
\begin{align}\label{eqn:j0_spa_eqn}
    F(\omega, y) &= \frac{\omega}{i} \exp\left[i\omega \left( \frac{y^2}{2} - \phi_{\mathrm{min}}(y)\right) \right]\nonumber\\
    &\quad \times \int_{0}^{\infty} rdr \exp \left[ i\omega \left( \frac{r^2}{2} -\ln(r) \right) \right] J_0 (\omega r y).
\end{align}

We use the following relation \cite{gradshteyn2014table} for the integral,
\begin{align}
    \int_{0}^{\infty} r^{\mu} e^{-\alpha r^2} J_{\nu}(\beta r) dr &= \frac{\beta^{\nu} \Gamma \left( \frac{\nu + \mu + 1}{2} \right)}{2^{\nu + 1} \alpha^{\frac{\nu + \mu + 1}{2}} \Gamma (\nu + 1)} \nonumber \\& \quad \times {}_1F_1 \left( \frac{\nu + \mu + 1}{2}, \nu + 1; \frac{-\beta^2}{4\alpha}\right).
\end{align}
In this general form of the integral we identify the relevant quantities as follows:
$\mu = 1 - i\omega$, $\nu = 0$, $\alpha = -\frac{i\omega}{2}$ and $\beta = \omega y$ and use the identity $\left( i a \right)^{ia} = e^{\frac{a\pi}{2}} \cdot e^{ia\log(|a|)}$, where $a = - \frac{\omega}{2}$. The lensing amplification factor then takes the form:
\begin{align}
    F(\omega, y) &= \exp\left[ \frac{\omega \pi}{4} + \frac{i\omega}{2} \left[ \log\left(\frac{\omega}{2} \right) -2\phi_{\mathrm{min}}(y) \right]\right] \nonumber\\ &\quad \times \Gamma\left( 1 - \frac{i\omega}{2} \right) {}_1F_1 \left( 1 - \frac{i\omega}{2}, 1 ; \frac{-i\omega y^2}{2} \right) e^{\frac{i\omega y^2}{2}}.
\end{align}
Using Kummer's transformation for Hypergeometric functions \cite{gradshteyn2014table} ${}_1F_1 (a, b ; z) = e^{z}  {}_1F_1(b-a, b ; -z)$, we obtain, 
 \begin{align}
    F(\omega, y) &= \exp\left[ \frac{\omega \pi}{4} + \frac{i\omega}{2} \left[ \log\left(\frac{\omega}{2} \right) -2\phi_{\mathrm{min}}(y) \right]\right] \nonumber\\ &\quad \times \Gamma\left( 1 - \frac{i\omega}{2} \right) {}_1F_1 \left( \frac{i\omega}{2}, 1 ; \frac{i\omega y^2}{2} \right).
 \end{align}

\subsection{Geometric optics approximation for $F(\omega, y)$}\label{subsec:geom_ff_derivation}

Consider the radial integral appearing in Eq.~(\ref{eqn:j0_spa_eqn}). We write it as,
\begin{equation}
\mathcal{G}(\omega, y)
= \int_{0}^{\infty} r \,
\exp\!\left[i \omega \!\left( \frac{r^2}{2} - \ln r \right)\right]
J_0(\omega r y)\, dr .
\end{equation}
For large arguments, the Bessel function admits the following asymptotic form:
\begin{equation}
J_0(x) \sim 
\sqrt{\frac{2}{\pi x}}
\cos\!\left(x-\frac{\pi}{4}\right),
\qquad x\to\infty .
\end{equation}
We can use this form in the large $\omega$ limit. Then the radial integral becomes,
\begin{align}
\mathcal{G}(\omega,y)
&\sim 
\begin{aligned}[t]
\int_{0}^{\infty} r \,
e^{\,i\omega\left(\frac{r^2}{2}-\ln r\right)}
\sqrt{\frac{2}{\pi \omega r y}}
\cos\!\left(\omega r y-\frac{\pi}{4}\right) dr
\\[6pt]
= \frac{1}{\sqrt{2\pi\omega y}}
\Bigg[
e^{i\pi/4}\!\int_{0}^{\infty}\!\sqrt{r}\,
e^{\,i\omega\left(\frac{r^2}{2}-\ln r-ry\right)} dr
\\
\qquad
+\,e^{-i\pi/4}\!\int_{0}^{\infty}\!\sqrt{r}\,
e^{\,i\omega\left(\frac{r^2}{2}-\ln r+ry\right)} dr
\Bigg]
\end{aligned}
\end{align}

It is convenient to introduce the integrals
\begin{subequations}
\begin{align}
\mathcal{G}_{+}(\omega,y)
&= \int_{0}^{\infty} \sqrt{r}\,
e^{\,i\omega \phi_{+}(r,y)}\, dr, \\
\mathcal{G}_{-}(\omega,y)
&= \int_{0}^{\infty} \sqrt{r}\,
e^{\,i\omega \phi_{-}(r,y)}\, dr, 
\end{align}
\end{subequations}

where, 
\begin{subequations}
\begin{align}
\phi_{+}(r,y)
&= \frac{r^2}{2}-\ln(r)-ry, \\[6pt]
\phi_{-}(r,y)
&= \frac{r^2}{2}-\ln(r)+ry .
\end{align}
\end{subequations}

The radial integral can then be written as,
\begin{equation}
\mathcal{G}(\omega,y)
= \frac{1}{\sqrt{2\pi\omega y}}
\left[
e^{i\pi/4}\mathcal{G}_{+}(\omega,y)
+
e^{-i\pi/4}\mathcal{G}_{-}(\omega,y)
\right].
\end{equation}
The stationary points of the phase functions $\phi_\pm$ follow from:
\begin{align}\label{eq:lens_eq}
\phi'_{\pm}(r) = r-\frac{1}{r}\mp y &= 0, \nonumber\\
\Rightarrow\quad r^2 \mp ry -1 &= 0 .
\end{align}

The physically relevant roots ($r>0$) are,
\begin{equation}
r_{\pm}=\frac{\pm y+\sqrt{y^2+4}}{2}.
\end{equation}
The second derivatives at the stationary points are:
\begin{equation}
\phi''_{\pm}(r_{\pm})=1+\frac{1}{r_{\pm}^2}.
\end{equation}
Using the stationary phase approximation (SPA), the integrals
$\mathcal{G}_{\pm}$ take the large-$\omega$ form,
\begin{align}
\mathcal{G}_{\pm}(\omega,y)
&\simeq
\sqrt{\frac{2\pi}{\omega |\phi''_{\pm}(r_{\pm})|}}
\nonumber \\ &\quad \times \sqrt{r_{\pm}}\,
\exp\!\left[
i\omega \phi_{\pm}(r_{\pm})
+ i\,\frac{\pi}{4}\,\mathrm{sign}\!\big(\phi''_{\pm}(r_{\pm})\big)
\right].
\end{align}

For $y>0$, $\mathrm{sign}(\phi''_{\pm})=+1$, and therefore
\begin{equation}
\begin{aligned}
\mathcal{G}(\omega,y)
&= \frac{1}{\sqrt{2\pi\omega y}}
\Bigg[
e^{i\pi/4}\sqrt{\frac{2\pi r_{+}}{\omega|\phi''_{+}(r_{+})|}}
e^{\,i\omega\phi_{+}(r_{+})+i\pi/4} \\[4pt]
&\qquad +
e^{-i\pi/4}\sqrt{\frac{2\pi r_{-}}{\omega|\phi''_{-}(r_{-})|}}
e^{\,i\omega\phi_{-}(r_{-})+i\pi/4}
\Bigg] \\[6pt]
&= \frac{1}{\omega}
\left[
i\,A_{+}(y)\,e^{i\omega\phi_{+}(r_{+})}
+ A_{-}(y)\,e^{i\omega\phi_{-}(r_{-})}
\right].
\end{aligned}
\end{equation}
where
\begin{equation}
A_{\pm}(y)=\sqrt{\frac{r_{\pm}}{y\,|\phi''_{\pm}(r_{\pm})|}} .
\end{equation}
Using the lens equation (\ref{eq:lens_eq}), which in our notation takes the form $yr_{\pm} = \pm(r^{2}_{\pm} - 1)$, the  prefactors $A_{\pm}$ can be simplified as follows: 
\bea
A_{\pm}^2 (y) &=& \frac{r_{\pm}^3}{y(1+r_{\pm}^2)} = \frac{r_{\pm}^4}{y\,r_{\pm}(1+r_{\pm}^2)} \,, \no \\
&=& \pm \frac{1}{1-1/r_{\pm}^{4}} \equiv \pm \mu_{\pm} \,.
\eea
Note that $\mu_{-} < 0$. It can be shown that,
\begin{equation}
    \mu_{\pm} = \frac{1}{2} \pm \frac{y^2 + 2}{2y\sqrt{y^2 + 4}}.
\end{equation}

Thus, the leading order behaviour of $F(\omega, y)$ is,
\begin{equation}
    \begin{aligned}
            F(\omega, y) &= \frac{\omega}{i} \exp\left[i\omega \left( \frac{y^2}{2} - \phi_{\mathrm{min}}(y)\right) \right] \mathcal{G}(\omega, y) \\
             &= \frac{\omega}{i} \exp\left[i\omega \left( \frac{y^2}{2} - \phi_{\mathrm{min}}(y)\right) \right] \nonumber \\ &\quad \times\frac{i}{\omega} ~ \left[ \sqrt{|\mu_{+}|}\mathrm{e}^{i \omega \phi_{+}(r_{+})} - i \sqrt{|\mu_{-}|}\mathrm{e}^{i \omega \phi_{-}(r_{-})} \right]
    \end{aligned}
\end{equation}
It is easily checked that, 
\begin{equation}
        \frac{y^2}{2} - \phi_m(y) = -\phi_{+}(r_{+}) \,,
\end{equation}
which then brings the magnification factor to a convenient form. We then have,
\begin{equation}\label{eq:geom_ff}
    \begin{aligned}
        F(\omega, y) &= \mathrm{e}^{-i\omega \phi_{+}(r_{+})} \left[ \sqrt{|\mu_{+}|} \mathrm{e}^{i\omega \phi_{+}(r_{+})} - i\sqrt{|\mu_{-}|}~\mathrm{e}^{i\omega \phi_{-}(r_{-})} \right]\\
        &= \sqrt{|\mu_{+}|} -i \sqrt{|\mu_{-}|}\mathrm{e}^{2\pi i f t_{\mathrm{d}}(M_{\mathrm{Lz}}, y)} \,,
    \end{aligned}
\end{equation}
where,
\begin{equation}
    t_\mathrm{d}(M_{\mathrm{Lz}}, y) = \frac{8\pi \mathrm{G} M_{Lz}}{\mathrm{c}^3}\left[ \frac{y\sqrt{y^2 + 4}}{2} + \mathrm{ln}\left( \frac{y + \sqrt{y^2 + 4}}{-y + \sqrt{y^2 + 4}} \right) \right] \,.
\end{equation}

As an aside, we also present a potentially useful transformation of  Eq. (\ref{eq:geom_ff}) written in $\zeta$ coordinates, where 
\begin{equation}
    \sinh(\zeta) = \frac{y}{2}.
\end{equation}
The $t_{\mathrm{d}}$ and $\mu_{\pm}$ in these new coordinates turn out to be, 
\begin{align}
    t_\mathrm{d}(M_{\mathrm{Lz}}, \zeta) &= \frac{8\pi G M_{Lz}}{c^3} \left[ 2\zeta + \sinh(2\zeta) \right] \,, \\
    \mu_{\pm}(\zeta) &= \frac{1}{2}\left[ 1 \pm \coth(2\zeta) \right].
\end{align}
The form of $t_\mathrm{d}$ is suggestive of the free fall time for a particle falling into a Schwarzschild black hole. However, here, instead of circular functions, hyperbolic functions appear in the expression because the orbit is unbound. 

\subsection{Derivation of the phase metric}
\label{appendix:phase_metric}

The unlensed waveform can be written in the following form:
\begin{equation}
    h_{\rm UL} (f;\lambda) = A(\lambda)f^{-7/6} \, e^{i\phi (f;\lambda)} ,
\end{equation}
where $\lambda$ denotes the set of parameters which are required to be searched by using the CBC template bank. The amplitude is slowly varying, so it is only the phase that matters in the placement of templates - hence the phase metric.
When the waveform is lensed with lensing parameters $\lambda'$ the unlensed waveform is multiplied by the magnification factor $F(f, \lambda')$ which is of the form,
\be
F(f, \lambda') = A' (\lambda')~e^{i \phi' (f; \lambda')} \,,
\ee
where $A'$ weakly depends on $f$ except for an insignificant interval of time-delays between the two images. We therefore omit the $f$ dependence from the amplitude $A'$ and write it as a function of $\lambda'$ only. So here as well, the phase decides the template placement. 
For a detector characterized by a one-sided noise power spectral density $S_h(f)$, the noise-weighted inner product which is complex (different from the usual real one, and appropriate for the phase metric) is defined as, 
\begin{equation}
    (a|b) = \int_{f_{\mathrm{low}}}^{f_{\mathrm{high}}}
    \frac{a^*(f)\, b(f)}{S_h(f)}\, \mathrm{d}f .
\end{equation}
The lensed waveform, which we just denote by $h$ in order to avoid clutter, must have norm unity and so,
\begin{align}
    (h|h) &= \int_{f_{\mathrm{low}}}^{f_{\mathrm{high}}}
    \frac{h^*(f;\lambda, \lambda')\, h(f;\lambda, \lambda')}{S_h(f)}\, \mathrm{d}f \\
    &= A^2(\lambda) A'^2 (\lambda') 
    \int_{f_{\mathrm{low}}}^{f_{\mathrm{high}}}
    \frac{f^{-7/3}}{S_h(f)}\, \mathrm{d}f \\
    &\equiv A^2(\lambda) A'^2 (\lambda') \, \mathcal{N} \equiv 1 \,,
\end{align}
where $\mathcal{N}$ is the normalisation factor. Let us denote the CBC and lensing parameters by $\Lambda = \{\lambda, \lambda' \}$ and the total phase by $\Phi = \phi + \phi'$. The normalized waveform is therefore
\begin{equation}
    \hat{h}(f;\Lambda) =
    \frac{h(f;\Lambda)}{\sqrt{(h|h)}} =
    \frac{f^{-7/6}\, e^{i\Phi(f;\Lambda)}}{\sqrt{\mathcal{N}}}.
\end{equation}
The inner product between a normalized signal $\hat{h}(f; \Lambda)$ and a perturbed one $\hat{h}(f; \Lambda + \Delta \Lambda)$ is given by,
\bea
\label{eq:match_cbc}
\mathcal{M}(\Lambda,\Delta\Lambda)
&=& \vert \bigl(\,\hat h(f;\Lambda)\mid \hat h(f;\Lambda + \Delta \Lambda)\,\bigr) \vert \, \no \\
&=& \left \vert \frac{1}{\mathcal{N}}
   \int_{f_{\rm low}}^{f_{\rm high}}
     \frac{e^{i \left[ \Phi(f; \Lambda + \Delta \Lambda) - \Phi(f; \Lambda) \right]}}{f^{7/3} S_h (f)} \, df
  \right \vert \,.
\eea
As defined in the main text, we write down the weighted average of a quantity $Q$ as:
\begin{equation}
    \langle Q \rangle = \frac{1}{\mathcal{N}} \times \int_{f_{\mathrm{low}}}^{f_\mathrm{high}} \frac{f^{-7/3}}{S_h (f)} Q(f) \, df.
\end{equation}
The exponential $e^{i \Delta \Phi}$, where $\Delta \Phi \equiv \Phi(\Lambda + \Delta \Lambda) - \Phi(\Lambda)$ can be approximated upto $\mathcal{O}(\Delta \Phi^2)$, and so we obtain,
\bea
    \mathcal{M}(\Lambda,\Delta\Lambda) 
    &=&  \left\lvert \langle e^{i \Delta \Phi} \rangle \right \rvert \, \no \\    
    &\approx& 
  \left\lvert
    \frac{1}{\mathcal{N}} \int_{f_{\mathrm{low}}}^{f_{\mathrm{hi gh}}}
      ~\frac{1 + i\Delta \Phi - \Delta \Phi^2 / 2}{f^{7/3} S_h (f)} \,~ df
  \right\rvert.
\label{eq:match_phase}  
\eea

Upto the first order, $\Delta \Phi \approx \frac{\partial \Phi}{\partial \Lambda^{\alpha}} \Delta \Lambda^{\alpha} \equiv \Phi_{\alpha} \Delta \Lambda^{\alpha} \break \implies ~ (\Delta \Phi)^2 \approx \Phi_{\alpha} \Phi_{\beta} \Delta \Lambda^{\alpha} \Delta \Lambda^{\beta}$

Expanding the match to second order in
$\Delta\Lambda^\alpha$ gives
\begin{align}
\mathcal{M}(\Lambda,\Delta\Lambda)
&\approx 
\left|
\langle
1+i\Phi_\alpha\Delta\Lambda^\alpha
-\frac{1}{2}\Phi_\alpha\Phi_\beta
\Delta\Lambda^\alpha\Delta\Lambda^\beta \rangle
\right|.
\end{align}

From the linearity of averaging operation,
\begin{equation}
\mathcal{M} \approx
\left|
1+i\,\langle \Phi_\alpha \rangle \Delta\Lambda^\alpha
-\frac{1}{2} \langle \Phi_\alpha\Phi_\beta \rangle
\Delta\Lambda^\alpha \Delta\Lambda^\beta
\right|.
\end{equation}

Writing out the magnitude of a complex quantity, we have,
\begin{align}
\mathcal{M}
&\approx 
\Big[
\left(1-\tfrac12
\langle \Phi_\alpha \Phi_\beta \rangle
\Delta\Lambda^\alpha \Delta\Lambda^\beta\right)^2 \nonumber \\
&\quad \qquad+ \langle \Phi_\alpha \rangle \langle \Phi_\beta \rangle
\Delta \Lambda^\alpha \Delta\Lambda^\beta
\Big]^{1/2}.
\end{align}

Keeping terms up to $\mathcal{O}(\Delta\lambda^2)$,
\begin{equation}
\mathcal{M}(\Lambda,\Delta\Lambda)
\approx
1-\frac12
\left[\langle \Phi_\alpha \Phi_\beta \rangle
-\langle \Phi_\alpha \rangle \langle \Phi_\beta \rangle
\right]
\Delta\Lambda^\alpha \Delta\Lambda^\beta
\end{equation}

Relating the match to the metric as, 
\begin{equation}
    \mathcal{M}(\Lambda,\Delta\Lambda) = 1 - g_{\alpha \beta} \Delta \Lambda^{\alpha} \Delta \Lambda^{\beta},
\end{equation}

we obtain,

\begin{equation}
    \boxed{g_{\alpha \beta} = \frac{1}{2} \left[\langle \Phi_{\alpha} \Phi_{\beta} \rangle - \langle \Phi_{\alpha} \rangle \langle \Phi_{\beta} \rangle \right]}.
\end{equation}
This completes the derivation of the phase metric for the full parameter space.


\bibliography{references}

@article{chan2025detectability,
  title={Detectability of lensed gravitational waves in matched-filtering searches},
  author={Chan, Juno CL and Seo, Eungwang and Li, Alvin KY and Fong, Heather and Ezquiaga, Jose M},
  journal={Physical Review D},
  volume={111},
  number={8},
  pages={084019},
  year={2025},
  publisher={APS}
}

@article{gholap2025chi,
  title={$\chi$ 2 statistic for the identification of strongly lensed gravitational waves from compact binary coalescences},
  author={Gholap, Sudhir and Soni, Kanchan and Kapadia, Shasvath J and Dhurandhar, Sanjeev},
  journal={Physical Review D},
  volume={112},
  number={12},
  pages={124074},
  year={2025},
  publisher={APS}
}

@article{mishra2024unveiling,
  title={Unveiling microlensing biases in testing general relativity with gravitational waves},
  author={Mishra, Anuj and Krishnendu, NV and Ganguly, Apratim},
  journal={Physical Review D},
  volume={110},
  number={8},
  pages={084009},
  year={2024},
  publisher={APS}
}

@article{ligo2026gwtc,
  title={GWTC-5.0: Observations from the Second Part of the Fourth LIGO-Virgo-KAGRA Observing Run and Updates to the Gravitational-Wave Transient Catalog},
  author={LIGO Scientific Collaboration and Virgo Collaboration and KAGRA Collaboration and others},
  journal={arXiv preprint arXiv:2605.27225},
  year={2026}
}

@article{kopty2026optimal,
  title={Optimal cross-correlation technique to search for strongly lensed gravitational waves},
  author={Kopty, Anirban and Mitra, Sanjit and More, Anupreeta},
  journal={arXiv preprint arXiv:2601.22138},
  year={2026}
}

@article{chakraborty2024glance,
  title={GLANCE--gravitational lensing authenticator using non-modelled cross-correlation exploration of gravitational wave signals},
  author={Chakraborty, Aniruddha and Mukherjee, Suvodip},
  journal={Monthly Notices of the Royal Astronomical Society},
  volume={532},
  number={4},
  pages={4842--4863},
  year={2024},
  publisher={Oxford University Press}
}

@article{goyal2021rapid,
  title={Rapid identification of strongly lensed gravitational-wave events with machine learning},
  author={Goyal, Srashti and Kapadia, Shasvath J and Ajith, Parameswaran},
  journal={Physical Review D},
  volume={104},
  number={12},
  pages={124057},
  year={2021},
  publisher={APS}
}

@book{SchneiderEhlersFalco1992,
  author    = {Peter Schneider and J{\"u}rgen Ehlers and Emilio E. Falco},
  title     = {Gravitational Lenses},
  publisher = {Springer-Verlag},
  address   = {Berlin},
  year      = {1992},
  doi       = {10.1007/978-3-662-03758-4}
}

@article{takahashi2003wave,
  title={Wave effects in the gravitational lensing of gravitational waves from chirping binaries},
  author={Takahashi, Ryuichi and Nakamura, Takashi},
  journal={The Astrophysical Journal},
  volume={595},
  number={2},
  pages={1039--1051},
  year={2003}
}

@article{LIGOScientific:2014pky,
    author = "Aasi, J. and others",
    collaboration = "LIGO Scientific",
    title = "{Advanced LIGO}",
    eprint = "1411.4547",
    archivePrefix = "arXiv",
    primaryClass = "gr-qc",
    doi = "10.1088/0264-9381/32/7/074001",
    journal = "Class. Quant. Grav.",
    volume = "32",
    pages = "074001",
    year = "2015"
}

@article{VIRGO:2014yos,
    author = "Acernese, F. and others",
    collaboration = "VIRGO",
    title = "{Advanced Virgo: a second-generation interferometric gravitational wave detector}",
    eprint = "1408.3978",
    archivePrefix = "arXiv",
    primaryClass = "gr-qc",
    doi = "10.1088/0264-9381/32/2/024001",
    journal = "Class. Quant. Grav.",
    volume = "32",
    number = "2",
    pages = "024001",
    year = "2015"
}

@article{KAGRA:2020tym,
    author = "Akutsu, T. and others",
    collaboration = "KAGRA",
    title = "{Overview of KAGRA: Detector design and construction history}",
    eprint = "2005.05574",
    archivePrefix = "arXiv",
    primaryClass = "physics.ins-det",
    doi = "10.1093/ptep/ptaa125",
    journal = "PTEP",
    volume = "2021",
    number = "5",
    pages = "05A101",
    year = "2021"
}

@article{cannon2021gstlal,
  title={GstLAL: A software framework for gravitational wave discovery},
  author={Cannon, Kipp and Caudill, Sarah and Chan, Chiwai and Cousins, Bryce and Creighton, Jolien DE and Ewing, Becca and Fong, Heather and Godwin, Patrick and Hanna, Chad and Hooper, Shaun and others},
  journal={SoftwareX},
  volume={14},
  pages={100680},
  year={2021},
  publisher={Elsevier}
}

@article{allene2025mbta,
  title={The MBTA pipeline for detecting compact binary coalescences in the fourth LIGO-Virgo-KAGRA observing run},
  author={All{\'e}n{\'e}, Christopher and Aubin, Florian and Bentara, In{\`e}s and Buskulic, Damir and Guidi, Gianluca M and Juste, Vincent and Lethuillier, Morgan and Marion, Fr{\'e}d{\'e}rique and Mobilia, Lorenzo and Mours, Beno{\^\i}t and others},
  journal={Classical and Quantum Gravity},
  volume={42},
  number={10},
  pages={105009},
  year={2025},
  publisher={IOP Publishing}
}

@article{dal2021real,
  title={Real-time search for compact binary mergers in Advanced LIGO and Virgo's third observing run using PyCBC live},
  author={Dal Canton, Tito and Nitz, Alexander H and Gadre, Bhooshan and Cabourn Davies, Gareth S and Villa-Ortega, Veronica and Dent, Thomas and Harry, Ian and Xiao, Liting},
  journal={The Astrophysical Journal},
  volume={923},
  number={2},
  pages={254},
  year={2021},
  publisher={The American Astronomical Society}
}

@article{chu2022spiir,
  title={SPIIR online coherent pipeline to search for gravitational waves from compact binary coalescences},
  author={Chu, Qi and Kovalam, Manoj and Wen, Linqing and Slaven-Blair, Teresa and Bosveld, Joel and Chen, Yanbei and Clearwater, Patrick and Codoreanu, Alex and Du, Zhihui and Guo, Xiangyu and others},
  journal={Physical Review D},
  volume={105},
  number={2},
  pages={024023},
  year={2022},
  publisher={APS}
}

@article{Owen:1995tm,
    author = "Owen, Benjamin J.",
    title = "{Search templates for gravitational waves from inspiraling binaries: Choice of template spacing}",
    eprint = "gr-qc/9511032",
    archivePrefix = "arXiv",
    doi = "10.1103/PhysRevD.53.6749",
    journal = "Phys. Rev. D",
    volume = "53",
    pages = "6749--6761",
    year = "1996"
}

@article{BSD-1,
  title = {Choice of filters for the detection of gravitational waves from coalescing binaries},
  author = {Sathyaprakash, B. S. and Dhurandhar, S. V.},
  journal = {Phys. Rev. D},
  volume = {44},
  issue = {12},
  pages = {3819--3834},
  numpages = {0},
  year = {1991},
  month = {Dec},
  publisher = {American Physical Society},
  doi = {10.1103/PhysRevD.44.3819},
  url = {https://link.aps.org/doi/10.1103/PhysRevD.44.3819}
}

@article{BSD-2,
  title = {Choice of filters for the detection of gravitational waves from coalescing binaries. II. Detection in colored noise},
  author = {Dhurandhar, S. V. and Sathyaprakash, B. S.},
  journal = {Phys. Rev. D},
  volume = {49},
  issue = {4},
  pages = {1707--1722},
  numpages = {0},
  year = {1994},
  month = {Feb},
  publisher = {American Physical Society},
  doi = {10.1103/PhysRevD.49.1707},
  url = {https://link.aps.org/doi/10.1103/PhysRevD.49.1707}
}

@article{BSO,
  title = {Matched filtering of gravitational waves from inspiraling compact binaries: Computational cost and template placement},
  author = {Owen, Benjamin J. and Sathyaprakash, B. S.},
  journal = {Phys. Rev. D},
  volume = {60},
  issue = {2},
  pages = {022002},
  numpages = {12},
  year = {1999},
  month = {Jun},
  publisher = {American Physical Society},
  doi = {10.1103/PhysRevD.60.022002},
  url = {https://link.aps.org/doi/10.1103/PhysRevD.60.022002}
}

@article{DalCanton2015,
  title = {Impact of precession on aligned-spin searches for neutron-star--black-hole binaries},
  author = {Dal Canton, Tito and Lundgren, Andrew P. and Nielsen, Alex B.},
  journal = {Phys. Rev. D},
  volume = {91},
  issue = {6},
  pages = {062010},
  numpages = {11},
  year = {2015},
  month = {Mar},
  publisher = {American Physical Society},
  doi = {10.1103/PhysRevD.91.062010},
  url = {https://link.aps.org/doi/10.1103/PhysRevD.91.062010}
}

@article{Bustillo2016,
  title = {Impact of gravitational radiation higher order modes on single aligned-spin gravitational wave searches for binary black holes},
  author = {Calder\'on Bustillo, Juan and Husa, Sascha and Sintes, Alicia M. and P\"urrer, Michael},
  journal = {Phys. Rev. D},
  volume = {93},
  issue = {8},
  pages = {084019},
  numpages = {12},
  year = {2016},
  month = {Apr},
  publisher = {American Physical Society},
  doi = {10.1103/PhysRevD.93.084019},
  url = {https://link.aps.org/doi/10.1103/PhysRevD.93.084019}
}

@article{Bustillo2021,
  title = {Confusing Head-On Collisions with Precessing Intermediate-Mass Binary Black Hole Mergers},
  author = {Bustillo, Juan Calder\'on and Sanchis-Gual, Nicolas and Torres-Forn\'e, Alejandro and Font, Jos\'e A.},
  journal = {Phys. Rev. Lett.},
  volume = {126},
  issue = {20},
  pages = {201101},
  numpages = {6},
  year = {2021},
  month = {May},
  publisher = {American Physical Society},
  doi = {10.1103/PhysRevLett.126.201101},
  url = {https://link.aps.org/doi/10.1103/PhysRevLett.126.201101}
}

@article{Dietrich2019,
  title = {Matter imprints in waveform models for neutron star binaries: Tidal and self-spin effects},
  author = {Dietrich, Tim and Khan, Sebastian and Dudi, Reetika and Kapadia, Shasvath J. and Kumar, Prayush and Nagar, Alessandro and Ohme, Frank and Pannarale, Francesco and Samajdar, Anuradha and Bernuzzi, Sebastiano and Carullo, Gregorio and Del Pozzo, Walter and Haney, Maria and Markakis, Charalampos and P\"urrer, Michael and Riemenschneider, Gunnar and Setyawati, Yoshinta Eka and Tsang, Ka Wa and Van Den Broeck, Chris},
  journal = {Phys. Rev. D},
  volume = {99},
  issue = {2},
  pages = {024029},
  numpages = {23},
  year = {2019},
  month = {Jan},
  publisher = {American Physical Society},
  doi = {10.1103/PhysRevD.99.024029},
  url = {https://link.aps.org/doi/10.1103/PhysRevD.99.024029}
}

@article{Ramos-Buades2020,
  title = {Impact of eccentricity on the gravitational-wave searches for binary black holes: High mass case},
  author = {Ramos-Buades, Antoni and Tiwari, Shubhanshu and Haney, Maria and Husa, Sascha},
  journal = {Phys. Rev. D},
  volume = {102},
  issue = {4},
  pages = {043005},
  numpages = {13},
  year = {2020},
  month = {Aug},
  publisher = {American Physical Society},
  doi = {10.1103/PhysRevD.102.043005},
  url = {https://link.aps.org/doi/10.1103/PhysRevD.102.043005}
}

@article{Phukon2025,
  title = {Geometric template bank for the detection of spinning low-mass compact binaries with moderate orbital eccentricity},
  author = {Phukon, Khun Sang and Schmidt, Patricia and Pratten, Geraint},
  journal = {Phys. Rev. D},
  volume = {111},
  issue = {4},
  pages = {043040},
  numpages = {28},
  year = {2025},
  month = {Feb},
  publisher = {American Physical Society},
  doi = {10.1103/PhysRevD.111.043040},
  url = {https://link.aps.org/doi/10.1103/PhysRevD.111.043040}
}

@article{Chia2024,
  title = {In pursuit of Love numbers: First templated search for compact objects with large tidal deformabilities in the LIGO-Virgo data},
  author = {Chia, Horng Sheng and Edwards, Thomas D. P. and Wadekar, Digvijay and Zimmerman, Aaron and Olsen, Seth and Roulet, Javier and Venumadhav, Tejaswi and Zackay, Barak and Zaldarriaga, Matias},
  journal = {Phys. Rev. D},
  volume = {110},
  issue = {6},
  pages = {063007},
  numpages = {31},
  year = {2024},
  month = {Sep},
  publisher = {American Physical Society},
  doi = {10.1103/PhysRevD.110.063007},
  url = {https://link.aps.org/doi/10.1103/PhysRevD.110.063007}
}

@article{Mehta:2025jiq,
    author = "Mehta, Ajit Kumar and Wadekar, Digvijay and Roulet, Javier and Anantpurkar, Isha and Venumadhav, Tejaswi and Mushkin, Jonathan and Zackay, Barak and Zaldarriaga, Matias and Islam, Tousif",
    title = "{Significant increase in sensitive volume of a gravitational wave search upon including higher harmonics}",
    eprint = "2501.17939",
    archivePrefix = "arXiv",
    primaryClass = "gr-qc",
    month = "1",
    year = "2025",
    journal = ""
}

@article{Ng2018,
  title = {Precise LIGO lensing rate predictions for binary black holes},
  author = {Ng, Ken K. Y. and Wong, Kaze W. K. and Broadhurst, Tom and Li, Tjonnie G. F.},
  journal = {Phys. Rev. D},
  volume = {97},
  issue = {2},
  pages = {023012},
  numpages = {6},
  year = {2018},
  month = {Jan},
  publisher = {American Physical Society},
  doi = {10.1103/PhysRevD.97.023012},
  url = {https://link.aps.org/doi/10.1103/PhysRevD.97.023012}
}

@article{li2018gravitational,
  title={Gravitational lensing of gravitational waves: A statistical perspective},
  author={Li, Shun-Sheng and Mao, Shude and Zhao, Yuetong and Lu, Youjun},
  journal={Monthly Notices of the Royal Astronomical Society},
  volume={476},
  number={2},
  pages={2220--2229},
  year={2018},
  publisher={Oxford University Press}
}

@article{oguri2018effect,
  title={Effect of gravitational lensing on the distribution of gravitational waves from distant binary black hole mergers},
  author={Oguri, Masamune},
  journal={Monthly Notices of the Royal Astronomical Society},
  volume={480},
  number={3},
  pages={3842--3855},
  year={2018},
  publisher={Oxford University Press}
}

@article{smith2018if,
  title={What if LIGO’s gravitational wave detections are strongly lensed by massive galaxy clusters?},
  author={Smith, Graham P and Jauzac, Mathilde and Veitch, John and Farr, Will M and Massey, Richard and Richard, Johan},
  journal={Monthly Notices of the Royal Astronomical Society},
  volume={475},
  number={3},
  pages={3823--3828},
  year={2018},
  publisher={Oxford University Press}
}

@article{Smith:2019qsv,
    author = "Smith, Graham P. and Robertson, Andrew and Bianconi, Matteo and Jauzac, Mathilde",
    collaboration = "LSST",
    title = "{Discovery of Strongly-lensed Gravitational Waves - Implications for the LSST Observing Strategy}",
    eprint = "1902.05140",
    archivePrefix = "arXiv",
    primaryClass = "astro-ph.HE",
    month = "2",
    year = "2019",
    journal = ""
}

@article{smith2020massively,
  title={Massively parallel Bayesian inference for transient gravitational-wave astronomy},
  author={Smith, Rory JE and Ashton, Gregory and Vajpeyi, Avi and Talbot, Colm},
  journal={Monthly Notices of the Royal Astronomical Society},
  volume={498},
  number={3},
  pages={4492--4502},
  year={2020},
  publisher={Oxford University Press}
}

@article{robertson2020does,
  title={What does strong gravitational lensing? The mass and redshift distribution of high-magnification lenses},
  author={Robertson, Andrew and Smith, Graham P and Massey, Richard and Eke, Vincent and Jauzac, Mathilde and Bianconi, Matteo and Ryczanowski, Dan},
  journal={Monthly Notices of the Royal Astronomical Society},
  volume={495},
  number={4},
  pages={3727--3739},
  year={2020},
  publisher={Oxford University Press}
}

@article{ryczanowski2020building,
  title={On building a cluster watchlist for identifying strongly lensed supernovae, gravitational waves and kilonovae},
  author={Ryczanowski, Dan and Smith, Graham P and Bianconi, Matteo and Massey, Richard and Robertson, Andrew and Jauzac, Mathilde},
  journal={Monthly Notices of the Royal Astronomical Society},
  volume={495},
  number={2},
  pages={1666--1671},
  year={2020},
  publisher={Oxford University Press}
}

@article{Dai:2017huk,
    author = "Dai, Liang and Venumadhav, Tejaswi",
    title = "{On the waveforms of gravitationally lensed gravitational waves}",
    eprint = "1702.04724",
    archivePrefix = "arXiv",
    primaryClass = "gr-qc",
    month = "2",
    year = "2017",
    journal = ""
}

@article{LIGOScientific:2025cwb,
    author = "Abac, A. G. and others",
    collaboration = "LIGO Scientific, VIRGO, KAGRA",
    title = "{GWTC-4.0: Searches for Gravitational-Wave Lensing Signatures}",
    eprint = "2512.16347",
    archivePrefix = "arXiv",
    primaryClass = "gr-qc",
    reportNumber = "LIGO-P2500419",
    month = "12",
    year = "2025",
    journal = ""
}

@article{Haris:2018vmn,
    author = "Haris, K. and Mehta, Ajit Kumar and Kumar, Sumit and Venumadhav, Tejaswi and Ajith, Parameswaran",
    title = "{Identifying strongly lensed gravitational wave signals from binary black hole mergers}",
    eprint = "1807.07062",
    archivePrefix = "arXiv",
    primaryClass = "gr-qc",
    reportNumber = "LIGO- P1800155",
    month = "7",
    year = "2018",
    journal = ""
}

@inproceedings{Janquart:2022wxc,
    author = "Janquart, Justin and Hannuksela, Otto A. and Haris, K. and Van den Broeck, Chris",
    title = "{GOLUM: A fast and precise methodology to search for, and analyze, strongly lensed gravitational-wave events}",
    booktitle = "{56th Rencontres de Moriond on Gravitation}",
    eprint = "2203.06444",
    archivePrefix = "arXiv",
    primaryClass = "gr-qc",
    month = "3",
    year = "2022"
}

@article{Janquart:2023osz,
    author = "Janquart, Justin and Haris, K. and Hannuksela, Otto A. and Van Den Broeck, Chris",
    title = "{The return of GOLUM: improving distributed joint parameter estimation for strongly lensed gravitational waves}",
    eprint = "2304.12148",
    archivePrefix = "arXiv",
    primaryClass = "gr-qc",
    doi = "10.1093/mnras/stad2838",
    journal = "Mon. Not. Roy. Astron. Soc.",
    volume = "526",
    number = "2",
    pages = "3088--3098",
    year = "2023"
}

@article{lo2023,
  title = {Bayesian statistical framework for identifying strongly lensed gravitational-wave signals},
  author = {Lo, Rico K. L. and Maga\~na Hernandez, Ignacio},
  journal = {Phys. Rev. D},
  volume = {107},
  issue = {12},
  pages = {123015},
  numpages = {26},
  year = {2023},
  month = {Jun},
  publisher = {American Physical Society},
  doi = {10.1103/PhysRevD.107.123015},
  url = {https://link.aps.org/doi/10.1103/PhysRevD.107.123015}
}

@article{barsode2025fast,
  title={Fast and efficient bayesian method to search for strongly lensed gravitational waves},
  author={Barsode, Ankur and Goyal, Srashti and Ajith, Parameswaran},
  journal={The Astrophysical Journal},
  volume={980},
  number={2},
  pages={258},
  year={2025},
  publisher={The American Astronomical Society}
}

@article{Ezquiaga2023,
  title = {Identifying strongly lensed gravitational waves through their phase consistency},
  author = {Ezquiaga, Jose Mar\'{\i}a and Hu, Wayne and Lo, Rico K. L.},
  journal = {Phys. Rev. D},
  volume = {108},
  issue = {10},
  pages = {103520},
  numpages = {23},
  year = {2023},
  month = {Nov},
  publisher = {American Physical Society},
  doi = {10.1103/PhysRevD.108.103520},
  url = {https://link.aps.org/doi/10.1103/PhysRevD.108.103520}
}

@article{Campailla2026,
  title = {Machine learning assisted parameter-space searches for lensed gravitational waves},
  author = {Campailla, Giulia and Raveri, Marco and Hu, Wayne and Ezquiaga, Jose Mar\'{\i}a},
  journal = {Phys. Rev. D},
  volume = {113},
  issue = {8},
  pages = {083510},
  numpages = {28},
  year = {2026},
  month = {Apr},
  publisher = {American Physical Society},
  doi = {10.1103/j87w-hpvx},
  url = {https://link.aps.org/doi/10.1103/j87w-hpvx}
}

@article{li2023,
  title = {Targeted subthreshold search for strongly lensed gravitational-wave events},
  author = {Li, Alvin K. Y. and Lo, Rico K. L. and Sachdev, Surabhi and Chan, Juno C. L. and Lin, E. T. and Li, Tjonnie G. F. and Weinstein, Alan J.},
  journal = {Phys. Rev. D},
  volume = {107},
  issue = {12},
  pages = {123014},
  numpages = {14},
  year = {2023},
  month = {Jun},
  publisher = {American Physical Society},
  doi = {10.1103/PhysRevD.107.123014},
  url = {https://link.aps.org/doi/10.1103/PhysRevD.107.123014}
}

@article{LIGOScientific:2023bwz,
    author = "Abbott, R. and others",
    collaboration = "LIGO Scientific, KAGRA, VIRGO",
    title = "{Search for Gravitational-lensing Signatures in the Full Third Observing Run of the LIGO{\textendash}Virgo Network}",
    eprint = "2304.08393",
    archivePrefix = "arXiv",
    primaryClass = "gr-qc",
    reportNumber = "LIGO-P2200031",
    doi = "10.3847/1538-4357/ad3e83",
    journal = "Astrophys. J.",
    volume = "970",
    number = "2",
    pages = "191",
    year = "2024"
}

@article{LIGOScientific:2021izm,
    author = "Abbott, R. and others",
    collaboration = "LIGO Scientific, VIRGO",
    title = "{Search for Lensing Signatures in the Gravitational-Wave Observations from the First Half of LIGO{\textendash}Virgo{\textquoteright}s Third Observing Run}",
    eprint = "2105.06384",
    archivePrefix = "arXiv",
    primaryClass = "gr-qc",
    reportNumber = "LIGO-P2000400",
    doi = "10.3847/1538-4357/ac23db",
    journal = "Astrophys. J.",
    volume = "923",
    number = "1",
    pages = "14",
    year = "2021"
}

@article{cheung2021stellar,
  title={Stellar-mass microlensing of gravitational waves},
  author={Cheung, Mark HY and Gais, Joseph and Hannuksela, Otto A and Li, Tjonnie GF},
  journal={Monthly Notices of the Royal Astronomical Society},
  volume={503},
  number={3},
  pages={3326--3336},
  year={2021},
  publisher={Oxford University Press}
}

@article{shan2025interference,
  title={An interference-based method for the detection of strongly lensed gravitational waves},
  author={Shan, Xikai and Hu, Bin and Chen, Xuechun and Cai, Rong-Gen},
  journal={Nature Astronomy},
  volume={9},
  number={6},
  pages={916--924},
  year={2025},
  publisher={Nature Publishing Group UK London}
}

@article{Basak:2021ten,
    author = "Basak, S. and Ganguly, A. and Haris, K. and Kapadia, S. and Mehta, A. K. and Ajith, P.",
    title = "{Constraints on Compact Dark Matter from Gravitational Wave Microlensing}",
    eprint = "2109.06456",
    archivePrefix = "arXiv",
    primaryClass = "gr-qc",
    reportNumber = "LIGO-P2100321",
    doi = "10.3847/2041-8213/ac4dfa",
    journal = "Astrophys. J.",
    volume = "926",
    number = "2",
    pages = "L28",
    year = "2022"
}

@article{Prabhu:2025elp,
    author = "Prabhu, Gopalkrishna and Deka, Uddeepta and Chakraborty, Sumanta and Kapadia, Shasvath J.",
    title = "{Probing the spin of compact objects with gravitational microlensing of gravitational waves}",
    eprint = "2512.18707",
    archivePrefix = "arXiv",
    primaryClass = "gr-qc",
    month = "12",
    year = "2025",
    journal = ""
}

@article{Deka:2024ecp,
    author = "Deka, Uddeepta and Chakraborty, Sumanta and Kapadia, Shasvath J. and Shaikh, Md Arif and Ajith, Parameswaran",
    title = "{Probing the charge of compact objects with gravitational microlensing of gravitational waves}",
    eprint = "2401.06553",
    archivePrefix = "arXiv",
    primaryClass = "gr-qc",
    doi = "10.1103/PhysRevD.111.064028",
    journal = "Phys. Rev. D",
    volume = "111",
    number = "6",
    pages = "064028",
    year = "2025"
}

@article{Seo:2021psp,
    author = "Seo, Eungwang and Hannuksela, Otto A. and Li, Tjonnie G. F.",
    title = "{Improving Detection of Gravitational-wave Microlensing Using Repeated Signals Induced by Strong Lensing}",
    eprint = "2110.03308",
    archivePrefix = "arXiv",
    primaryClass = "astro-ph.HE",
    doi = "10.3847/1538-4357/ac6dea",
    journal = "Astrophys. J.",
    volume = "932",
    number = "1",
    pages = "50",
    year = "2022"
}

@article{Lai2018,
  title = {Discovering intermediate-mass black hole lenses through gravitational wave lensing},
  author = {Lai, Kwun-Hang and Hannuksela, Otto A. and Herrera-Mart\'{\i}n, Antonio and Diego, Jose M. and Broadhurst, Tom and Li, Tjonnie G. F.},
  journal = {Phys. Rev. D},
  volume = {98},
  issue = {8},
  pages = {083005},
  numpages = {7},
  year = {2018},
  month = {Oct},
  publisher = {American Physical Society},
  doi = {10.1103/PhysRevD.98.083005},
  url = {https://link.aps.org/doi/10.1103/PhysRevD.98.083005}
}

@ARTICLE{Deguchi1986,
       author = {{Deguchi}, S. and {Watson}, W.~D.},
        title = "{Diffraction in Gravitational Lensing for Compact Objects of Low Mass}",
      journal = {\apj},
         year = 1986,
        month = aug,
       volume = {307},
        pages = {30},
          doi = {10.1086/164389},
       adsurl = {https://ui.adsabs.harvard.edu/abs/1986ApJ...307...30D}
}

@article{Nakamura1998,
  title = {Gravitational Lensing of Gravitational Waves from Inspiraling Binaries by a Point Mass Lens},
  author = {Nakamura, Takahiro T.},
  journal = {Phys. Rev. Lett.},
  volume = {80},
  issue = {6},
  pages = {1138--1141},
  numpages = {0},
  year = {1998},
  month = {Feb},
  publisher = {American Physical Society},
  doi = {10.1103/PhysRevLett.80.1138},
  url = {https://link.aps.org/doi/10.1103/PhysRevLett.80.1138}
}

@article{Goyal2024,
  title = {Rapid method for preliminary identification of subthreshold strongly lensed counterparts to superthreshold gravitational-wave events},
  author = {Goyal, Srashti and Kapadia, Shasvath J. and Cudell, Jean-Ren\'e and Li, Alvin K. Y. and Chan, Juno C. L.},
  journal = {Phys. Rev. D},
  volume = {109},
  issue = {2},
  pages = {023028},
  numpages = {12},
  year = {2024},
  month = {Jan},
  publisher = {American Physical Society},
  doi = {10.1103/PhysRevD.109.023028},
  url = {https://link.aps.org/doi/10.1103/PhysRevD.109.023028}
}

@article{Ezquiaga2021,
  title = {Phase effects from strong gravitational lensing of gravitational waves},
  author = {Ezquiaga, Jose Mar\'{\i}a and Holz, Daniel E. and Hu, Wayne and Lagos, Macarena and Wald, Robert M.},
  journal = {Phys. Rev. D},
  volume = {103},
  issue = {6},
  pages = {064047},
  numpages = {28},
  year = {2021},
  month = {Mar},
  publisher = {American Physical Society},
  doi = {10.1103/PhysRevD.103.064047},
  url = {https://link.aps.org/doi/10.1103/PhysRevD.103.064047}
}

@article{anadsensvd2002,
doi = {10.1088/0264-9381/19/7/337},
url = {https://dx.doi.org/10.1088/0264-9381/19/7/337},
year = {2002},
month = {mar},
publisher = {},
volume = {19},
number = {7},
pages = {1507},
author = {Anand S Sengupta and  Sanjeev V Dhurandhar and  Albert Lazzarini and  Tom Prince},
title = {Extended hierarchical search (EHS) algorithm for detection of gravitational waves from inspiralling compact binaries},
journal = {Classical and Quantum Gravity}
}

@article{anandsensvd2003,
  title = {Faster implementation of the hierarchical search algorithm for detection of gravitational waves from inspiraling compact binaries},
  author = {Sengupta, Anand S. and Dhurandhar, Sanjeev and Lazzarini, Albert},
  journal = {Phys. Rev. D},
  volume = {67},
  issue = {8},
  pages = {082004},
  numpages = {14},
  year = {2003},
  month = {Apr},
  publisher = {American Physical Society},
  doi = {10.1103/PhysRevD.67.082004},
  url = {https://link.aps.org/doi/10.1103/PhysRevD.67.082004}
}

@article{mohantysvd96,
  title = {Hierarchical search strategy for the detection of gravitational waves from coalescing binaries},
  author = {Mohanty, S. D. and Dhurandhar, S. V.},
  journal = {Phys. Rev. D},
  volume = {54},
  issue = {12},
  pages = {7108--7128},
  numpages = {0},
  year = {1996},
  month = {Dec},
  publisher = {American Physical Society},
  doi = {10.1103/PhysRevD.54.7108},
  url = {https://link.aps.org/doi/10.1103/PhysRevD.54.7108}
}

@article{mohantysvd98,
  title = {Hierarchical search strategy for the detection of gravitational waves from coalescing binaries: Extension to post-Newtonian waveforms},
  author = {Mohanty, S. D.},
  journal = {Phys. Rev. D},
  volume = {57},
  issue = {2},
  pages = {630--658},
  numpages = {0},
  year = {1998},
  month = {Jan},
  publisher = {American Physical Society},
  doi = {10.1103/PhysRevD.57.630},
  url = {https://link.aps.org/doi/10.1103/PhysRevD.57.630}
}

@article{kanchan_hierarchical,
  title = {Hierarchical search for compact binary coalescences in the Advanced LIGO's first two observing runs},
  author = {Soni, Kanchan and Gadre, Bhooshan Uday and Mitra, Sanjit and Dhurandhar, Sanjeev},
  journal = {Phys. Rev. D},
  volume = {105},
  issue = {6},
  pages = {064005},
  numpages = {16},
  year = {2022},
  month = {Mar},
  publisher = {American Physical Society},
  doi = {10.1103/PhysRevD.105.064005},
  url = {https://link.aps.org/doi/10.1103/PhysRevD.105.064005}
}

@article{bhooshan_hierarchical,
  title = {Hierarchical search strategy for the efficient detection of gravitational waves from nonprecessing coalescing compact binaries with aligned-spins},
  author = {Gadre, Bhooshan and Mitra, Sanjit and Dhurandhar, Sanjeev},
  journal = {Phys. Rev. D},
  volume = {99},
  issue = {12},
  pages = {124035},
  numpages = {19},
  year = {2019},
  month = {Jun},
  publisher = {American Physical Society},
  doi = {10.1103/PhysRevD.99.124035},
  url = {https://link.aps.org/doi/10.1103/PhysRevD.99.124035}
}

@misc{CE,
      title={Cosmic Explorer: A Submission to the NSF MPSAC ngGW Subcommittee}, 
      author={Matthew Evans and Alessandra Corsi and Chaitanya Afle and Alena Ananyeva and K. G. Arun and Stefan Ballmer and Ananya Bandopadhyay and Lisa Barsotti and Masha Baryakhtar and Edo Berger and Emanuele Berti and Sylvia Biscoveanu and Ssohrab Borhanian and Floor Broekgaarden and Duncan A. Brown and Craig Cahillane and Lorna Campbell and Hsin-Yu Chen and Kathryne J. Daniel and Arnab Dhani and Jennifer C. Driggers and Anamaria Effler and Robert Eisenstein and Stephen Fairhurst and Jon Feicht and Peter Fritschel and Paul Fulda and Ish Gupta and Evan D. Hall and Giles Hammond and Otto A. Hannuksela and Hannah Hansen and Carl-Johan Haster and Keisi Kacanja and Brittany Kamai and Rahul Kashyap and Joey Shapiro Key and Sanika Khadkikar and Antonios Kontos and Kevin Kuns and Michael Landry and Philippe Landry and Brian Lantz and Tjonnie G. F. Li and Geoffrey Lovelace and Vuk Mandic and Georgia L. Mansell and Denys Martynov and Lee McCuller and Andrew L. Miller and Alexander Harvey Nitz and Benjamin J. Owen and Cristiano Palomba and Jocelyn Read and Hemantakumar Phurailatpam and Sanjay Reddy and Jonathan Richardson and Jameson Rollins and Joseph D. Romano and Bangalore S. Sathyaprakash and Robert Schofield and David H. Shoemaker and Daniel Sigg and Divya Singh and Bram Slagmolen and Piper Sledge and Joshua Smith and Marcelle Soares-Santos and Amber Strunk and Ling Sun and David Tanner and Lieke A. C. van Son and Salvatore Vitale and Benno Willke and Hiro Yamamoto and Michael Zucker},
      year={2023},
      eprint={2306.13745},
      archivePrefix={arXiv},
      primaryClass={astro-ph.IM},
      url={https://arxiv.org/abs/2306.13745}, 
}

@article{ET,
doi = {10.1088/0264-9381/27/19/194002},
url = {https://doi.org/10.1088/0264-9381/27/19/194002},
year = {2010},
month = {sep},
publisher = {},
volume = {27},
number = {19},
pages = {194002},
author = {Punturo, M and Abernathy, M and Acernese, F and Allen, B and Andersson, N and Arun, K and Barone, F and Barr, B and Barsuglia, M and Beker, M and Beveridge, N and Birindelli, S and Bose, S and Bosi, L and Braccini, S and Bradaschia, C and Bulik, T and Calloni, E and Cella, G and Mottin, E Chassande and Chelkowski, S and Chincarini, A and Clark, J and Coccia, E and Colacino, C and Colas, J and Cumming, A and Cunningham, L and Cuoco, E and Danilishin, S and Danzmann, K and De Luca, G and De Salvo, R and Dent, T and De Rosa, R and Di Fiore, L and Di Virgilio, A and Doets, M and Fafone, V and Falferi, P and Flaminio, R and Franc, J and Frasconi, F and Freise, A and Fulda, P and Gair, J and Gemme, G and Gennai, A and Giazotto, A and Glampedakis, K and Granata, M and Grote, H and Guidi, G and Hammond, G and Hannam, M and Harms, J and Heinert, D and Hendry, M and Heng, I and Hennes, E and Hild, S and Hough, J and Husa, S and Huttner, S and Jones, G and Khalili, F and Kokeyama, K and Kokkotas, K and Krishnan, B and Lorenzini, M and Lück, H and Majorana, E and Mandel, I and Mandic, V and Martin, I and Michel, C and Minenkov, Y and Morgado, N and Mosca, S and Mours, B and Müller–Ebhardt, H and Murray, P and Nawrodt, R and Nelson, J and Oshaughnessy, R and Ott, C D and Palomba, C and Paoli, A and Parguez, G and Pasqualetti, A and Passaquieti, R and Passuello, D and Pinard, L and Poggiani, R and Popolizio, P and Prato, M and Puppo, P and Rabeling, D and Rapagnani, P and Read, J and Regimbau, T and Rehbein, H and Reid, S and Rezzolla, L and Ricci, F and Richard, F and Rocchi, A and Rowan, S and Rüdiger, A and Sassolas, B and Sathyaprakash, B and Schnabel, R and Schwarz, C and Seidel, P and Sintes, A and Somiya, K and Speirits, F and Strain, K and Strigin, S and Sutton, P and Tarabrin, S and Thüring, A and van den Brand, J and van Leewen, C and van Veggel, M and van den Broeck, C and Vecchio, A and Veitch, J and Vetrano, F and Vicere, A and Vyatchanin, S and Willke, B and Woan, G and Wolfango, P and Yamamoto, K},
title = {The Einstein Telescope: a third-generation gravitational wave observatory},
journal = {Classical and Quantum Gravity}
}

@software{pycbc,
  author       = {Alex Nitz and
                  Ian Harry and
                  Duncan Brown and
                  Christopher M. Biwer and
                  Josh Willis and
                  Tito Dal Canton and
                  Collin Capano and
                  Thomas Dent and
                  Larne Pekowsky and
                  Andrew R. Williamson and
                  Soumi De and
                  Miriam Cabero and
                  Bernd Machenschalk and
                  Duncan Macleod and
                  Prayush Kumar and
                  Steven Reyes and
                  Francesco Pannarale and
                  Gareth S Cabourn Davies and
                  dfinstad and
                  Sumit Kumar and
                  Márton Tápai and
                  Leo Singer and
                  Sebastian Khan and
                  Stephen Fairhurst and
                  Alex Nielsen and
                  Shashwat Singh and
                  Thomas Massinger and
                  Koustav Chandra and
                  Shasvath and
                  Veronica-villa},
  title        = {gwastro/pycbc: v2.0.4 release of PyCBC},
  month        = jun,
  year         = 2022,
  publisher    = {Zenodo},
  version      = {v2.0.4},
  doi          = {10.5281/zenodo.6646669},
  url          = {https://doi.org/10.5281/zenodo.6646669}
}

@article{vanderWalt:2011bqk,
    author = "van der Walt, Stéfan and Colbert, S. Chris and Varoquaux, Gaël",
    archivePrefix = "arXiv",
    doi = "10.1109/MCSE.2011.37",
    eprint = "1102.1523",
    journal = "Comput. Sci. Eng.",
    number = "2",
    pages = "22--30",
    primaryClass = "cs.MS",
    title = "{The NumPy Array: A Structure for Efficient Numerical Computation}",
    volume = "13",
    year = "2011"
}

@article{Virtanen:2019joe,
    author = "Virtanen, Pauli and others",
    archivePrefix = "arXiv",
    doi = "10.1038/s41592-019-0686-2",
    eprint = "1907.10121",
    journal = "Nature Meth.",
    primaryClass = "cs.MS",
    title = "{SciPy 1.0--Fundamental Algorithms for Scientific Computing in Python}",
    year = "2020"
}

@Article{Hunter:2007,
  Author    = {Hunter, J. D.},
  Title     = {Matplotlib: A 2D graphics environment},
  Journal   = {Computing in Science \& Engineering},
  Volume    = {9},
  Number    = {3},
  Pages     = {90--95},
  publisher = {IEEE COMPUTER SOC},
  doi       = {10.1109/MCSE.2007.55},
  year      = 2007
}

@article{2013A&A...558A..33A,
       title={Astropy: A community Python package for astronomy},
   volume={558},
   ISSN={1432-0746},
   url={http://dx.doi.org/10.1051/0004-6361/201322068},
   DOI={10.1051/0004-6361/201322068},
   journal={Astronomy \& amp; Astrophysics},
   publisher={EDP Sciences},
   author={Robitaille, Thomas P. and others},
   year={2013},
   month=sep, pages={A33} 
}

@article{2018AJ....156..123A,
       title={The Astropy Project: Building an Open-science Project and Status of the v2.0 Core Package*},
   volume={156},
   ISSN={1538-3881},
   url={http://dx.doi.org/10.3847/1538-3881/aabc4f},
   DOI={10.3847/1538-3881/aabc4f},
   number={3},
   journal={The Astronomical Journal},
   publisher={American Astronomical Society},
   author={Price-Whelan, A. M. and others},
   year={2018},
   month=aug, pages={123} }

@inproceedings{jupyter,
       booktitle = {Positioning and Power in Academic Publishing: Players, Agents and Agendas},
          editor = {Fernando Loizides and Birgit Scmidt},
           title = {Jupyter Notebooks - a publishing format for reproducible computational workflows},
          author = {Thomas Kluyver and Benjamin Ragan-Kelley and Fernando P{\'e}rez and Brian Granger and Matthias Bussonnier and Jonathan Frederic and Kyle Kelley and Jessica Hamrick and Jason Grout and Sylvain Corlay and Paul Ivanov and Dami{\'a}n Avila and Safia Abdalla and Carol Willing and  Jupyter development team},
       publisher = {IOS Press},
         address = {Netherlands},
            year = {2016},
           pages = {87--90},
             url = {https://eprints.soton.ac.uk/403913/}
}

@misc{gwmat,
  author       = {Anuj Mishra},
  title        = {{GWMAT: Gravitational Wave Microlensing Analysis Tools}},
  year         = {2025},
  howpublished = {\url{https://git.ligo.org/anuj.mishra/gwmat}},
  note         = {Accessed: 2026-06-26}
}

@article{husa2016frequency,
  title={Frequency-domain gravitational waves from nonprecessing black-hole binaries. I. New numerical waveforms and anatomy of the signal},
  author={Husa, Sascha and Khan, Sebastian and Hannam, Mark and P{\"u}rrer, Michael and Ohme, Frank and Forteza, Xisco Jim{\'e}nez and Boh{\'e}, Alejandro},
  journal={Physical Review D},
  volume={93},
  number={4},
  pages={044006},
  year={2016},
  publisher={APS}
}

@article{khan2016frequency,
  title={Frequency-domain gravitational waves from nonprecessing black-hole binaries. II. A phenomenological model for the advanced detector era},
  author={Khan, Sebastian and Husa, Sascha and Hannam, Mark and Ohme, Frank and P{\"u}rrer, Michael and Forteza, Xisco Jim{\'e}nez and Boh{\'e}, Alejandro},
  journal={Physical Review D},
  volume={93},
  number={4},
  pages={044007},
  year={2016},
  publisher={APS}
}

@article{buonanno2009comparison,
  title={Comparison of post-Newtonian templates for compact binary inspiral signals<? format?> in gravitational-wave detectors},
  author={Buonanno, Alessandra and Iyer, Bala R and Ochsner, Evan and Pan, Yi and Sathyaprakash, Bangalore Suryanarayana},
  journal={Physical Review D—Particles, Fields, Gravitation, and Cosmology},
  volume={80},
  number={8},
  pages={084043},
  year={2009},
  publisher={APS}
}

@article{cokelaer2007gravitational,
  title={Gravitational waves from inspiralling compact binaries: hexagonal template placement and its efficiency in detecting physical signals},
  author={Cokelaer, Thomas},
  journal={Physical Review D—Particles, Fields, Gravitation, and Cosmology},
  volume={76},
  number={10},
  pages={102004},
  year={2007},
  publisher={APS}
}

@article{harry2009stochastic,
  title={Stochastic template placement algorithm for gravitational wave data analysis},
  author={Harry, Ian W and Allen, Bruce and Sathyaprakash, BS},
  journal={Physical Review D—Particles, Fields, Gravitation, and Cosmology},
  volume={80},
  number={10},
  pages={104014},
  year={2009},
  publisher={APS}
}

@article{roy2017hybrid,
  title={Hybrid geometric-random template-placement algorithm for gravitational wave searches from compact binary coalescences},
  author={Roy, Soumen and Sengupta, Anand S and Thakor, Nilay},
  journal={Physical Review D},
  volume={95},
  number={10},
  pages={104045},
  year={2017},
  publisher={APS}
}

@book{gradshteyn2014table,
  title={Table of integrals, series, and products},
  author={Gradshteyn, Izrail Solomonovich and Ryzhik, Iosif Moiseevich},
  year={2014},
  publisher={Academic press}
}

@article{seo2025residual,
  title={Residual test to search for microlensing signatures in strongly lensed gravitational wave signals},
  author={Seo, Eungwang and Shan, Xikai and Janquart, Justin and Hannuksela, Otto and Hendry, Martin and Hu, Bin},
  journal={The Astrophysical Journal},
  volume={988},
  number={2},
  pages={159},
  year={2025},
  publisher={The American Astronomical Society}
}

@article{chakraborty2025mu,
  title={$\mu$-GLANCE: A Novel Technique to Detect Chromatically and Achromatically Lensed Gravitational-wave Signals},
  author={Chakraborty, Aniruddha and Mukherjee, Suvodip},
  journal={The Astrophysical Journal},
  volume={984},
  number={2},
  pages={107},
  year={2025},
  publisher={The American Astronomical Society}
}

@article{chakraborty2025first,
  title={The First Model-independent Chromatic Microlensing Search: No Evidence in the Gravitational Wave Catalog of LIGO--Virgo--KAGRA},
  author={Chakraborty, Aniruddha and Mukherjee, Suvodip},
  journal={The Astrophysical Journal},
  volume={990},
  number={1},
  pages={68},
  year={2025},
  publisher={The American Astronomical Society}
}

@article{chakraborty2026first,
  title={The First Model-independent Upper Bound on Microlensing Signature of the Highest-mass Binary Black Hole Event GW231123},
  author={Chakraborty, Aniruddha and Mukherjee, Suvodip},
  journal={The Astrophysical Journal},
  volume={1003},
  number={1},
  pages={20},
  year={2026},
  publisher={The American Astronomical Society}
}

\end{document}